\documentclass[aps,prd,10pt,twocolumn,preprintnumbers,tightenlines,superscriptaddress,nofootinbib,showpacs,longbibliography,floatfix]
{revtex4-1}
\usepackage{amssymb,latexsym}
\usepackage{amsmath,amsbsy,bbm}
\usepackage{epsfig,bm,color}
\usepackage{graphicx,comment}
\begin{document}

\def\a{{\alpha}}
\def\b{{\beta}}
\def\d{{\delta}}
\def\D{{\Delta}}
\def\X{{\Xi}}
\def\e{{\varepsilon}}
\def\g{{\gamma}}
\def\G{{\Gamma}}
\def\k{{\kappa}}
\def\l{{\lambda}}
\def\L{{\Lambda}}
\def\m{{\mu}}
\def\n{{\nu}}
\def\o{{\omega}}
\def\O{{\Omega}}
\def\S{{\Sigma}}
\def\s{{\sigma}}
\def\th{{\theta}}

\def\ol#1{{\overline{#1}}}

\def\sumintq{\hat{\sum} \hskip-1.35em \int_{ \hskip-0.25em \underset{q}{\phantom{a}} } \, \, \,}
\def\sumintk{\hat{\sum} \hskip-1.35em \int_{ \hskip-0.25em \underset{k}{\phantom{a}} } \, \, \,}
\def\sumintp{\hat{\sum} \hskip-1.35em \int_{ \hskip-0.25em \underset{p}{\phantom{a}} } \, \, \,}

\def\cF{{\mathcal F}}
\def\cS{{\mathcal S}}
\def\cC{{\mathcal C}}
\def\cB{{\mathcal B}}
\def\cT{{\mathcal T}}
\def\cQ{{\mathcal Q}}
\def\cL{{\mathcal L}}
\def\cO{{\mathcal O}}
\def\cA{{\mathcal A}}
\def\cQ{{\mathcal Q}}
\def\cR{{\mathcal R}}
\def\cH{{\mathcal H}}
\def\cW{{\mathcal W}}
\def\cM{{\mathcal M}}
\def\cD{{\mathcal D}}
\def\cN{{\mathcal N}}
\def\cP{{\mathcal P}}
\def\cK{{\mathcal K}}
\def\cI{{\mathcal{I}}}
\def\cJ{{\mathcal{J}}}

\def\eqref#1{{(\ref{#1})}}

\preprint{INT-PUB-16-030}  
\preprint{KITP-16-141}

\title{Finite-Volume Corrections to Electromagnetic Masses for Larger-Than-Physical Electric Charges}

\author{Matthew E.~Matzelle}
\altaffiliation[Present address:]
{
Department of Physics, 
Northeastern University, 
Boston, MA 02115-5000, USA}        
\email[]{$\texttt{Matt.Matzelle@gmail.com}$}
\affiliation{
Department of Physics,
        The City College of New York,
        New York, NY 10031, USA}
\author{Brian~C.~Tiburzi}
\email[]{$\texttt{bctiburz@gmail.com}$}
\affiliation{
Department of Physics,
        The City College of New York,
        New York, NY 10031, USA}
\affiliation{
Graduate School and University Center,
        The City University of New York,
        New York, NY 10016, USA}
\affiliation{
RIKEN BNL Research Center,
        Brookhaven National Laboratory,
        Upton, NY 11973, USA}
\affiliation{
Institute for Nuclear Theory, 
	University of Washington, 
	Seattle, WA 98195-1560, USA}
\affiliation{
Kavli Institute for Theoretical Physics, 
	University of California, 
	Santa Barbara, CA 93106, USA}

\date{\today}


\begin{abstract}

The numerical value of the fine-structure constant generally leads to small isospin-breaking effects due to electromagnetism in QCD.  
This smallness complicates determining isospin breaking from lattice QCD computations that include electromagnetism. 
One solution to this problem consists of performing computations using larger-than-physical values of the electric charge, 
and subsequently extrapolating (or interpolating) to the physical value of the fine-structure constant.  
Motivated by recent lattice QCD + QED computations of electromagnetic masses employing this setup, 
we consider finite-volume effects arising from the use of larger-than-physical electric charges. 
A modified power-counting scheme, 
which is based on treating the fine-structure constant as larger than its physical value, 
is explored. 
Results for perturbative QED corrections, 
however, 
are surprising. 
Within the framework of non-relativistic QED,
multi-loop diagrams exhibit a momentum factorization property that produces exact cancellations. 
We determine that power-law finite-volume effects vanish at the leading two- and three-loop order, 
as well as the next-to-leading two-loop order. 
For larger-than-physical charges, 
we consequently expect no appreciable volume corrections beyond leading-order QED. 
\end{abstract}
\maketitle

\section{Introduction}                                                 %

The study of strong interactions from first principles lattice QCD computations continues to make dramatic progress. 
Precision computations of the spectrum of hadrons have progressed to the level of addressing isospin breaking, 
which arises from the difference in light quark masses, 
as well as the difference in their electric charges. 
The first study of QED effects on hadron masses was performed in 
Ref.~\cite{Duncan:1996xy}; 
and, 
in recent years,  
considerable advances have been achieved by various lattice QCD collaborations%
~\cite{Blum:2007cy,Basak:2008na,Blum:2010ym,Aoki:2012st,Ishikawa:2012ix,Borsanyi:2013lga,deDivitiis:2013xla}.
Most recently, 
the impressive lattice computation of isospin splittings of hadrons performed in 
Refs.~\cite{Borsanyi:2014jba,Fodor:2016bgu}
includes completely both sources of isospin violation, 
the latter is achieved through the inclusion of fully dynamical QED. 
There are further areas for which electromagnetic corrections become relevant and the computations appear feasible:
for example, 
charged-particle scattering at low energies%
~\cite{Beane:2014qha}, 
and the necessity of QED for high-precision determination of hadronic processes%
~\cite{Carrasco:2015xwa,Tantalo:2016vxk}. 
In all of these cases, 
our lack of quantitative understanding of electromagnetic interactions is at the femtoscale, 
where QCD is non-perturbative and the lattice method is the only known tool available for systematically controlled calculations.

One important aspect arising in lattice QCD + QED calculations is the systematic effect from the finite volume of the lattice. 
By necessity, 
lattice computations are performed in finite spacetime volumes, 
and the long-range nature of QED interactions is subject to modification leading to power-law finite-volume corrections to observable quantities. 
Extrapolations to infinite volume are required, 
and analytic techniques using effective field theory have proven useful. 
In particular, 
the theoretical framework of non-relativistic QED
(NRQED)%
~\cite{Caswell:1985ui}
applied to hadronic bound states, 
see, 
e.g., 
Ref.~\cite{Hill:2011wy}, 
has been utilized to compute finite-volume corrections to the electromagnetic masses of hadrons%
~\cite{Davoudi:2014qua,Fodor:2015pna,Lee:2015rua}. 
A salient feature of this approach is the characterization of the power-law finite-volume dependence of observables in terms of a few low-energy hadronic properties
(charge, charge radii, magnetic moments, \emph{et c}.). 
Versatility is another feature, 
because the NRQED approach can be adapted to differing formulations of finite volume QED, 
for example, 
the inclusion of a non-vanishing photon mass to regulate the infrared behavior%
~\cite{Endres:2015gda}, 
or the implementation of $C^*$-boundary conditions%
~\cite{Wiese:1991ku},
which has been detailed as a strictly local, gauge-invariant formulation of QED in finite volume%
~\cite{Lucini:2015hfa}.

In this work, 
we apply the NRQED framework to a different strategy for lattice QCD + QED computations. 
The smallness of isospin breaking effects due to QED can be overcome computationally by 
inflating the size of the fine-structure constant, 
$\alpha$,
and subsequently performing an extrapolation to its physical value. 
If one also combines this approach with computations in pure QCD, 
one can perform an interpolation to the physical value of the fine-structure constant. 
This strategy has been adopted, 
for example, 
in the recent lattice QCD + QED computations of
Refs.%
~\cite{Horsley:2015eaa,Horsley:2015vla}, 
which determined isospin splittings in the spectrum of hadrons. 
Our focus is to better understand the systematics of this approach;
thereby, 
we consider finite-volume effects for larger-than-physical values of the electric charge.

The organization of our investigation is as follows. 
Introducing a modified power-counting scheme in 
Sec.~\ref{s:0}, 
we discuss the higher-order corrections that are relevant for finite-volume electromagnetic masses for larger-than-physical values of the electric charge.  
Results from bare perturbation theory required for our calculation are also summarized.  
We perform the leading two-loop NRQED computation of the finite-volume corrections to electromagnetic masses in 
Sec.~\ref{s:HOT}.  
The computation including its renormalization is rather technical, 
however, 
the non-relativistic treatment leads to dramatic simplifications. 
At this order, 
all individual contributions contain non-analytic dependence on the volume;
however, 
the separation between long- and short-distance scales is maintained by an exact cancellation of these problematic terms. 
Consequently the leading two-loop volume correction vanishes. 
Consideration of further higher-order corrections is also taken up in 
Sec.~\ref{s:HOT}. 
We additionally compute the leading three-loop and next-to-leading two-loop contributions to the electromagnetic masses. 
Both calculations exhibit simplifications due to the non-relativistic framework. 
In particular, 
the multi-loop diagrams exhibit a non-trivial momentum factorization property that leads to cancellations. 
We determine that power-law finite-volume corrections vanish at these higher orders too. 
Consistency of the calculations is confirmed in the Appendices.  
In Appendix~\ref{s:A}, 
the Ward identity is verified to the leading two-loop order and next-to-leading one-loop order. 
The leading two-loop results are alternately obtained using renormalized perturbation theory in Appendix~\ref{s:B}, 
and confirmed by deriving the counterterms directly from bare perturbation theory. 
 A summary in Sec.~\ref{s:summy} concludes this work.

\section{Modified Power-Counting Scheme}
\label{s:0}

To reduce systematic uncertainties, 
precision study of isospin breaking in the hadron spectrum using larger-than-physical values of the electric charge might need to address what are ordinarily minute
QED corrections. 
These are proportional to 
$\alpha^2$ 
and beyond. 
Such corrections modify both the infinite-volume and finite-volume results; 
the latter can be computed in an effective field theory framework and is our focus throughout.

\subsection{NRQED}

A tool for the efficient computation of power-law finite-volume effects encountered in lattice 
QCD + QED 
studies is that of effective field theory. 
Specifically NRQED applied to hadronic bound states has been demonstrated to be efficacious%
~\cite{Davoudi:2014qua}. 
One of the criteria for applicability of NRQED requires that the finite-volume effects associated with QCD are significantly smaller than those of QED, 
so that infinite-volume couplings%
\footnote{
When this condition is not met,
correcting the NRQED couplings for exponentially small volume corrections is only part of the story. 
There are further operators allowed in finite volume NRQED that are not constrained by 
$SO(3)$
rotational invariance. 
The coefficients of all such operators, 
however, 
are exponentially suppressed in the infinite-volume limit. 
}
can be utilized in the NRQED effective action. 
We assume that this requirement has been met.

The NRQED framework has been employed for QCD bound states of spin
$j = 0$, 
$\frac{1}{2}$, 
and 
$1$, 
however, 
the leading corrections are universal in that they only depend on the charge of the hadron. 
This is exemplified by the NRQED Lagrangian density, 
which is organized in powers of the hadron's Compton wavelength. 
Writing out the lowest-order terms, 
we have
\begin{eqnarray}
\cL 
= 
- \frac{1}{4} F_{\mu \nu} F^{\mu \nu} 
+ 
\cL_0 + \frac{1}{2M} \cL_1 + \frac{1}{4 M^2} \cL_2 + \cdots
. \, \end{eqnarray}
The photon field is described using the electromagnetic field-strength tensor, 
$F_{\mu \nu} = \partial_\mu A_\nu - \partial_\nu A_\mu$. 
The leading term for the matter field, 
$\cL_0$, 
has the form
\begin{eqnarray}
\cL_0
&=&
\psi^\dagger
\, i D_0 
\psi
\label{eq:NRQED}
,\end{eqnarray}
which is independent of the hadron's 
spin, 
mass, 
and internal structure. 
The gauge-covariant derivative, 
$D_\mu = \partial_\mu + i Q e A_\mu$, 
contains the hadron's charge, 
$Q$, 
which appears in units of the magnitude of the electron's charge,
$e > 0$. 
Notice that because the charge-density interaction is spin independent, 
we can treat the spin components of the matter field 
$\psi$
implicitly. 
Beyond leading order, 
the hadron's mass becomes relevant. 
At next-to-leading order, 
the kinetic energy is contained in the term
\begin{eqnarray}
\cL_1 
= 
\psi^\dagger \bm{D}^2 \psi
\label{eq:KinE}
,\end{eqnarray}
which remains independent of the spin and internal structure of the hadron. 
Spin- and structure-dependent interactions are contained in 
$\cL_2$
and beyond. 
These terms are not required for the computations we perform.%
\footnote{
There is an additional structure-dependent term in 
$\cL_1$, 
which is the magnetic moment operator, 
$\psi^\dagger \bm{\sigma} \cdot \bm{B} \, \psi$. 
Because of its spin dependence, 
however, 
the first non-vanishing contributions to the electromagnetic mass require two insertions of this operator. 
These are consequently proportional to 
$M^{-2}$,
and are beyond the considerations in this work. 
}

Using this framework, 
we compute the electromagnetic mass in a cubic spatial volume of 
$L^3$. 
In a combined
QED and NRQED expansion, 
it is convenient to write contributions to the electromagnetic mass in the form
\begin{equation}
M(\alpha,L) = M 
+ \sum_{j >0} \, \sum_{k \geq 0} \Delta M^{(j,k)}
\label{eq:expansion1}
,\end{equation}
with the finite-volume corrections contained in
\begin{equation}
\Delta M^{(j,k)} / M
\sim 
\alpha^j / (ML)^k
\label{eq:expansion}
,\end{equation}
where the first index labels the perturbative QED order, 
while the second labels the NRQED order. 
The parameter
$M$
is taken to be the hadron's mass in infinite volume without QED. 
Notice that this scale has been integrated out of QED, and leads to the NRQED expansion. 
The infinite-volume contributions to the electromagnetic mass, 
namely those with 
$k=0$,
require non-perturbative QCD to determine, 
while the finite-volume contributions, 
$k>0$, 
arise from long-range physics and can be determined using the effective theory. 
The central goal is to utilize such results to perform infinite-volume extrapolations of lattice QCD computations.

Our considerations are motivated by the recent lattice QCD + QED calculation of 
Ref.~\cite{Horsley:2015eaa}, 
for example.
This calculation employs a value of 
$\ol \alpha = 0.099$
for the fine-structure constant, 
which is larger than the physical value by a factor of
$13.6$. 
The authors were able to perform an interpolation to the physical value of the fine-structure constant in the following way. 
Masses of mesons and baryons were calculated in QCD + QED, 
and parameterized to include all possible linear-order quark mass terms 
(expanded symmetrically about an 
$SU(3)$
flavor-symmetric point), 
and all terms linear in the fine-structure constant. 
Within uncertainties, 
the coefficients of quark-mass dependent terms were observed to be consistent with their pure QCD values, 
i.e.~the corresponding calculations performed with 
$\ol \alpha = 0$~\cite{Bietenholz:2011qq}. 
Up to small corrections, 
this guarantees that the masses depend linearly on the fine-structure constant. 
Results were then interpolated to the physical value of the electric charge by 
multiplying the electromagnetic corrections obtained by the factor 
$\alpha / \ol \alpha$. 
Any residual higher-order QED effects, 
however, 
will depend on the combination
$\alpha \, \ol \alpha$, 
at leading order in 
$\ol \a$. 
These contributions are then unphysically large and could potentially affect isospin splittings at the level of
$10 \%$. 
We are unable to address the impact on infinite-volume physics, 
which requires lattice QCD + QED computations at further values of 
$\ol \alpha$. 
Using NRQED, 
however, 
we can address the finite-volume corrections.

%
%
%
\begin{figure}
\resizebox{\linewidth}{!}{
\includegraphics{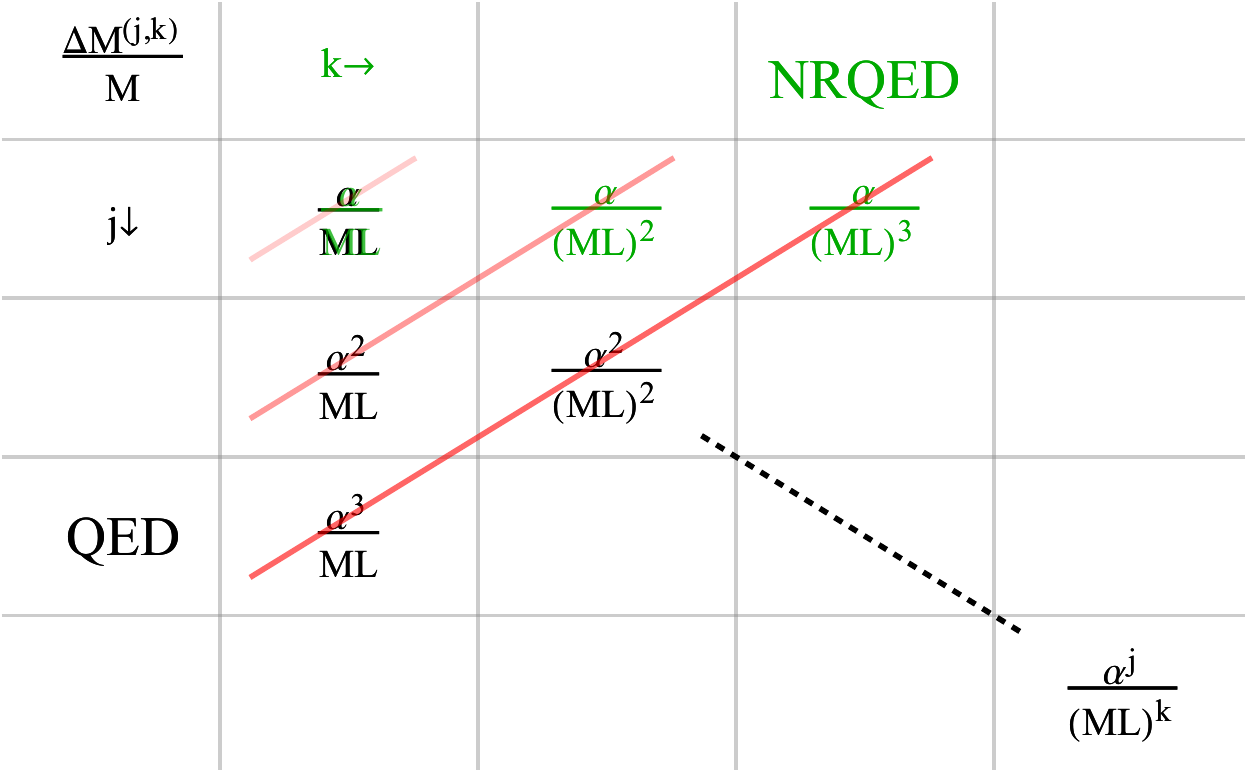}
}
\caption{
Depiction of power-counting schemes for finite-volume corrections in NRQED.
Contributions are grouped into rows and columns, 
where each row denotes a successive order in QED perturbation theory, 
and each column denotes a successive order in the NRQED expansion. 
At larger-than-physical values of the electric charge,
for example
$\alpha \sim 0.1$,  
the expansion should be organized in successive diagonals of this table. 
}
\label{f:PC}
\end{figure}
%
%
%

The power-counting scheme for finite-volume corrections employed by
Ref.~\cite{Davoudi:2014qua}  
is based on the physical value of the fine-structure constant.  
Consequently higher-order terms in the NRQED expansion are treated as much more important than higher-order terms in the perturbative 
QED expansion. 
With larger-than-physical values of the fine-structure constant, 
the perturbative QED corrections might become competitive with NRQED corrections. 
This situation is depicted in 
Fig.~\ref{f:PC}, 
where NRQED corrections and perturbative QED corrections are schematically compared. 
Unphysical values employed for the electric charge numerically satisfy the approximate relation
$\ol \alpha \sim \frac{1}{M L}$, 
which suggests that finite-volume corrections should be organized along the diagonals shown in the figure. 
Formally the reorganization of the expansion takes the form 
\begin{equation}
M(\alpha, L)
=
M
+ 
\Delta M (\alpha)
+
\sum_{i>1}
\overline{\Delta M} {}^{(i)}
,\end{equation}
where
\begin{equation}
\overline{\Delta M} {}^{(i)}
=
\sum_{j>0}^{i} \Delta M^{(j,\, i - j+1)}
,\end{equation}
and 
$\Delta M(\alpha)$ 
is used to represent the infinite-volume electromagnetic mass, 
i.e.
\begin{equation}
\Delta M(\alpha) = \sum_{j>0} \Delta M^{(j,0)}
.\end{equation} 
In this work, 
we compute the additional terms required for three orders,
$i = 1$--$3$,
of this modified power-counting scheme. 
The terms required are:
$\Delta M^{(2,1)}$, 
which is the leading two-loop finite-volume correction;
$\Delta M^{(3,1)}$, 
which is the leading three-loop finite-volume correction; 
and, 
finally,
$\Delta M^{(2,2)}$, 
which is the next-to-leading two-loop finite-volume correction. 
Surprisingly, 
we find all three of these corrections vanish.

\subsection{Bare Perturbation Theory}

Computations of finite-volume electromagnetic mass corrections in this work are performed using bare perturbation theory.
The pertinent details are described here. 
Additional details concerning wavefunction renormalization and verification of the Ward identity in this approach are given in 
Appendix~\ref{s:A}. 
Renormalized perturbation theory is alternately utilized for a subset of the computations in 
Appendix~\ref{s:B}.

For the general discussion presented in this section, 
it is convenient to explicitly treat just the QED expansion, 
and return to the NRQED expansion of each contribution only when required below. 
The hadron's mass expansion in QED is accordingly written in the simple form
\begin{equation}
\mathcal{M} 
\equiv
M(\alpha, L)  
-
M 
= 
\sum_{j>0} 
M^{(j)}
\label{eq:Mres}
,\end{equation}
where
$M^{(j)}$
has all contributions at order 
$\alpha^j$. 
This includes infinite-volume and finite-volume contributions, 
as well as all terms, 
in principle, 
occurring in the NRQED expansion. 
Due to the removal of the scale 
$M$
in NRQED, 
it is the residual mass
$\mathcal{M}$
that will be determined order by order.

In bare perturbation theory, 
ultraviolet divergences from loop corrections are rendered finite with the addition of counterterms. 
For our computation of the electromagnetic mass, 
we require the mass counterterm Lagrangian density
\begin{equation}
\cL_{c.t.}
=
- 
\sum_{j>0} 
\delta_M^{(j)}
\,
\psi^\dagger \psi 
,\end{equation}
where 
$j$ 
denotes the perturbative QED order,
namely 
$\delta_M^{(j)} \sim \alpha^j$.
Each counterterm 
has an additional expansion in NRQED, 
but this will be treated as needed below.%
\footnote{ 
An important aspect of NRQED is that the leading vacuum-polarization effect is proportional to 
$\mathcal{O}(\alpha / M^2)$. 
Because this is a correction to the photon propagator, 
the leading contribution to the hadron self-energy appears in the term
$\Delta M^{(2,3)}$,
whose computation is beyond the order that we work. 
For this reason, 
running of the QED coupling can be neglected throughout. 
}

The electromagnetic mass is determined from identifying the pole of the matter field's two-point correlation function. 
In NRQED, 
this correlation function has the general form
\begin{equation}
G(p_0)
=
\frac{i}{p_0 - \Sigma(p_0) + i \epsilon}
,\end{equation}
where the self-energy function 
$\Sigma(p_0)$
is determined from the sum of all one-particle irreducible contributions. 
The mass 
$\mathcal{M}$
is the solution to the equation
\begin{equation}
\mathcal{M} = \Sigma (\mathcal{M})
.\end{equation}
Expanding 
$p_0$
about this solution, 
we have the general behavior of the two-point function
\begin{equation}
G(p_0)
=
\frac{i Z}{p_0 - \mathcal{M} + i \epsilon}
+ 
\text{regular}
\label{eq:2pt}
,\end{equation}
with the wavefunction renormalization given by 
$Z^{-1} = 1 - \Sigma^\prime(p_0 = \mathcal{M})$, 
where the prime is used to denote differentiation with respect to 
$p_0$. 
Writing the sum of one-particle irreducible diagrams in the QED expansion, 
we have
\begin{equation}
\Sigma(p_0) = \sum_{j>0} \Sigma^{(j)}(p_0)
,\end{equation}
where 
$\Sigma^{(j)} \sim \alpha^j$. 
Comparing with the expansion of the residual mass in 
Eq.~\eqref{eq:Mres}, 
we can iteratively determine the solution for the pole position. 
The first two terms are given by familiar expressions
\begin{eqnarray}
M^{(1)} &=& \Sigma^{(1)}
,\notag \\
M^{(2)} 
&=& 
\Sigma^{(2)}
+ 
M^{(1)} \Sigma^{\prime(1)}
\label{eq:Mexp_1_2}
,\end{eqnarray}
while the third takes the form
\begin{eqnarray}
M^{(3)} 
&=& 
\Sigma^{(3)}
+ 
M^{(2)} \Sigma^{\prime(1)}
\notag \\
&& 
+ 
M^{(1)} \Sigma^{\prime(2)}
+ 
\frac{1}{2}  \left( M^{(1)}  \right)^2  \Sigma^{\prime \prime (1)}  
.\label{eq:Mexp_3}
\end{eqnarray}
When the argument of the self-energy function 
(or of its derivatives) has been omitted, 
we are implicitly referring to evaluation at 
$p_0 = 0$, 
for example, 
$\Sigma^{(1)} \equiv \Sigma^{(1)}(p_0 = 0)$, 
\emph{et c}. 
Computation of the terms appearing in these expressions comprises the remainder of this work. 
Connection of some of these terms with the wavefunction renormalization and verification of the Ward identity is carried out in Appendix~\ref{s:A}. 
Reorganization of the leading two-loop computation using renormalized perturbation theory appears in Appendix~\ref{s:B}.

\subsection{Requisite One-Loop Results}

\begin{figure}
\resizebox{0.8\linewidth}{!}{
\includegraphics{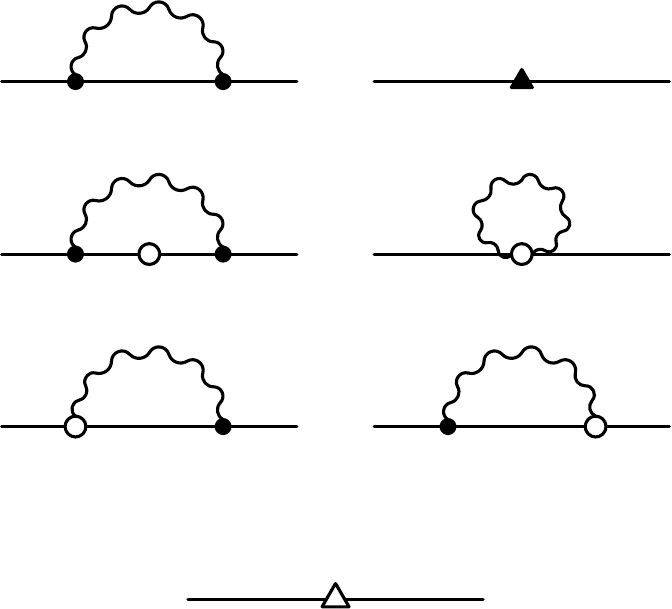}
}
\caption{
Graphical depiction of one-loop contributions to the electromagnetic mass of charged particles in NRQED. 
The top row shows the leading one-loop contribution along with the required counterterm, 
while the remaining rows show the next-to-leading one-loop contributions and their associated counterterm. 
In each diagram, 
the charged matter field is denoted by a straight line, 
while photons are denoted by wavy lines. 
The charge-density coupling between the photon and matter field appears as a filled circle. 
Open circles denote 
vertices 
formed from the kinetic-energy operator. 
Triangles are used to denote 
$\cO(\alpha)$
counterterms, 
with a solid triangle for the leading counterterm, 
and an open triangle for the next-to-leading counterterm. 
}
\label{f:oneloop}
\end{figure}

Before proceeding with the higher-order loop calculations, 
we report the results required from the one-loop computation. 
To work at the order desired, 
we require both the leading one-loop diagrams and next-to-leading one-loop diagrams in the NRQED expansion. 
Expanding the 
$\cO(\alpha)$
self-energy accordingly, 
we have
\begin{equation}
\Sigma^{(1)}(p_0) 
= 
\Sigma^{(1,1)}(p_0) 
+ 
\Sigma^{(1,2)}(p_0) 
+ 
\cdots
,\end{equation} 
where, 
in two-index form, 
the second index indicates the order in the NRQED expansion. 
The Feynman diagrams contributing to these two terms are depicted in 
Fig.~\ref{f:oneloop}.

%
%
%
%
%
%

\subsubsection{Leading One Loop}

At the leading one-loop order, 
we have the sunset diagram shown in Fig.~\ref{f:oneloop}. 
This loop contribution is 
\begin{eqnarray}
\Sigma_\text{1-loop}^{(1,1)} (p_0)
=
- i (Qe)^2
\,
\sumintq \frac{1}{q^2 + i \epsilon} \, \frac{1}{q_0 + p_0 + i \epsilon}
\label{eq:Sig_1_1_p0}
.\end{eqnarray}
We employ the shorthand notation
\begin{equation}
\sumintq
f(q)
=
\frac{1}{L^3}
\int_{- \infty}^{+ \infty} \frac{dq_0}{2\pi}
\sum_{\bm{q} = \frac{2 \pi \bm{n}}{L}  \neq \bm{0}} 
f(q)
,\end{equation}
for the integrals over the time-component of momentum
and sums over the quantized momentum modes permitted on a 
$3$-dimensional torus. 
The exclusion of spatial modes with
$\bm{n} = \bm{0}$
from all sums
arises from the treatment of the photon's zero momentum mode.%
\footnote{
These modes are typically excluded from QED gauge-field generation as a constraint, 
or can be dynamically generated, 
see~\cite{Gockeler:1991bu}. 
In either case, 
there is no contribution to the finite-volume action of the photon from zero modes, 
and the photon propagator appropriately reflects this.  
Finite-volume corrections to charged-hadron masses in NRQED are marginally different when these modes are included. 
The leading such difference takes the form of a kinetic energy term,  
$\frac{(Q e \bm{B})^2 }{2M} \sim \frac{\alpha}{M L^2}$, 
and this important contribution was accordingly subtracted from the results obtained in 
Ref.~\cite{Horsley:2015eaa}. 
There are additional corrections at higher orders. 
For example, 
insertion of this kinetic term inside the sunset diagram would seem to lead to a finite-volume correction of the form
$\sim \frac{\alpha^2}{ ML^2}$, 
however, 
this contribution is identically cancelled by wave-function renormalization. 
We find that a non-vanishing finite-volume correction does arise at one order higher in the NRQED expansion, 
$\sim \frac{\alpha^2}{ M^2 L^3}$, 
from one-loop diagrams formed with two insertions of the operator 
$\frac{Q e}{M} \psi^\dagger \bm{B} \cdot \bm{D} \, \psi$. 
In the notation of Eq.~\eqref{eq:expansion1}, 
this correction contributes to 
$\Delta M^{(2,3)}$, 
which is beyond the order considered in the present work. 
} 
The sum over momentum modes produces an ultraviolet divergent result, 
and the expression requires regularization. 
The infrared effect from the finite volume, 
however, 
is independent of regularization for a broad class of regulators including dimensional regularization and lattice regularization, 
see, e.g., Ref.~\cite{Hasenfratz:1989pk}. 
For this reason, 
we treat the regularization implicitly.

The necessary elements from the leading one-loop contribution appear in Eqs.~\eqref{eq:Mexp_1_2} and \eqref{eq:Mexp_3}. 
Carrying out the small 
$p_0$ 
expansion of Eq.~\eqref{eq:Sig_1_1_p0}, 
we find
\begin{eqnarray}
\Sigma_\text{1-loop}^{(1,1)}
&=&
\frac{(Q e)^2}{2}  \mathfrak{C}_2
,\notag \\
\Sigma_\text{1-loop}^{\prime(1,1)}
&=&
\frac{(Q e)^2}{2}  \mathfrak{C}_3
,\notag \\
\Sigma_\text{1-loop}^{\prime \prime(1,1)}
&=&
(Q e)^2 \mathfrak{C}_4
.\end{eqnarray} 
With the regularization treated as implicit, 
the $3$-dimensional momentum sums appearing above are all written in terms of 
\begin{equation}
\mathfrak{C}_j
=
\frac{1}{L^3}
\sum_{\bm{n} \neq \bm{0}}
\frac{1}{|\bm{q}|^j}
\label{eq:Cj}
,\end{equation}
where the periodic momentum modes are given by
$\bm{q} = \frac{2 \pi}{L} \bm{n}$. 
We will also make use of their infinite-volume limit, 
and according define the quantities
\begin{equation}
\hat{\mathfrak{C}}_j
=
\lim_{L \to \infty}
\mathfrak{C}_j
=
\int \frac{d \bm{q}}{(2 \pi)^3} \frac{1}{|\bm{q}|^j}
,\end{equation}
which are regularization dependent. 
In dimensionally regulated schemes, 
for example, 
we have
$\hat{\mathfrak{C}}_{j} \overset{DR}{=} 0$, 
for 
$j < 3$.

From the first of the results appearing above,  
we determine the leading-order mass counterterm 
\begin{equation}
\delta_M^{(1,1)} 
=
\Delta M^{(1,0)}
- 
\frac{(Qe)^2}{2}  \hat{\mathfrak{C}}_2
,\end{equation}
where
$\Delta M^{(1,0)}$
is the physical 
$\cO(\alpha)$
electromagnetic mass. 
Adding up the two contributions, 
we have
\begin{eqnarray}
M^{(1,1)}
&=&
\Sigma_\text{1-loop}^{(1,1)}
+ 
\delta_M^{(1,1)} 
=
\Delta M^{(1,0)} 
+ 
\frac{(Qe)^2}{2}  C_2
\label{eq:M_1_1}
.\end{eqnarray}
The first term is the infinite-volume result, 
and the second term is the finite-volume effect. 
The finite-volume effect has been written in terms of the ultraviolet finite difference%
\footnote{For the case 
$j=3$, 
there is an oddity.  
The subtraction of the corresponding infinite-volume integral,
$\int d^3 \bm{q} \,  |\bm{q}|^{-3}$, 
while curing the ultraviolet divergence,  
introduces an infrared divergence. 
Beyond one-loop order, 
such contributions appear
and could potentially lead to a dependence on the regularization scheme.  
We find, 
however, 
that all such contributions exactly cancel ensuring the scheme independence of the finite-volume effects. 
}
\begin{equation}
C_j 
= 
\mathfrak{C}_j - \hat{\mathfrak{C}}_j
,\end{equation}
which, 
in turn, 
can be written in terms of dimensionless shape coefficients, to which will make reference throughout. 
The relation is 
\begin{equation}
C_j = \frac{1}{L^3} \left(\frac{2 \pi}{L} \right)^{-j} c_j
,\end{equation}
with the shape coefficients defined by the sum of two terms
\begin{equation}
c_j = \frac{\pi^{\frac{j}{2}}}{\Gamma(\frac{j}{2})} \left( a_j + b_j \right)
,\end{equation} 
with these terms having the form
\begin{eqnarray}
a_j
&=& 
\int_0^1 ds
\left( 
s^{ - \frac{j}{2} - 1} 
+ 
s^{\frac{j}{2} - \frac{5}{2}}
\right)
\left[ \vartheta_3(0,e^{- \frac{\pi}{s}})^3 - 1 \right]
,\notag \\
b_j 
&=&
\frac{6}{j (j-3)}, 
\quad 
j \neq 3
.\end{eqnarray}
The function
$\vartheta_3(z,q)$
appearing above is a Jacobi elliptic-theta function. 
Upon utilizing the result 
$c_2 = \pi c_1$, 
we reproduce from Eq.~\eqref{eq:M_1_1} the leading finite-volume effect%
~\cite{Borsanyi:2014jba,Davoudi:2014qua}
\begin{equation}
\Delta M^{(1,1)} 
= 
\frac{(Qe)^2}{2}  C_2 
= 
\frac{Q^2 \alpha}{2 L}  c_1
.\end{equation}
Beyond reproducing this result, 
the determined mass counterterm and derivatives of the self-energy are required in the higher-order computations performed below.

\subsubsection{Next-To-Leading One Loop}

It is straightforward to extend the analysis to the next-to-leading one-loop contributions. 
The required one-particle irreducible diagrams are shown in 
Fig.~\ref{f:oneloop}. 
As a function of the temporal component of momentum,
$p_0$, 
we obtain the expression for the loop contributions
\begin{eqnarray}
\Sigma_\text{1-loop}^{(1,2)}
(p_0)
=
\frac{(Qe)^2}{2M} \,
\sumintq
\frac{- i}{q^2 + i \epsilon}
\left[
\frac{\bm{q}^2}{(q_0 + p_0 + i \epsilon)^2}
-
3
\right]
,
\notag\\
\end{eqnarray}
where we have dropped terms that trivially vanish. 
Notice the second term, 
which arises from the photon tadpole diagram, 
does not depend on
$p_0$. 
The required contributions to the electromagnetic mass are determined from the above expression to be
\begin{eqnarray}
\Sigma_\text{1-loop}^{(1,2)}
&=&
\frac{(Qe)^2}{2M} \mathfrak{C}_1
,\notag \\
\Sigma_\text{1-loop}^{\prime (1,2)}
&=&
-\frac{(Qe)^2}{2M} \mathfrak{C}_2
.\end{eqnarray}
Accordingly the next-to-leading order counterterm takes the form
\begin{equation}
\delta_M^{(1,2)}
= 
- \frac{(Qe)^2}{2M} \hat{\mathfrak{C}}_1
.\end{equation}
Finally, 
we recover the next-to-leading one-loop result for the finite-volume effect%
~\cite{Borsanyi:2014jba,Davoudi:2014qua}, 
namely
\begin{equation}
\Delta M^{(1,2)}
=
\frac{(Qe)^2}{2M} C_1
=
\frac{Q^2 \alpha}{M L^2}
c_1
.\end{equation}
Beyond this order, 
the mass counterterm and derivative of the self-energy function are required, 
and will be utilized below.

\section{Higher-Order Calculations}
\label{s:HOT}

To address the effect of larger-than-physical electric charges on the electromagnetic mass at finite volume, 
we perform higher-order QED calculations. 
With the modified power counting proposed above, 
there are three such calculations required. 
First we obtain the leading two-loop result, 
then the leading three-loop result, 
and finally the finite-volume correction appearing at next-to-leading two-loop order.

\subsection{Leading Two-Loop Calculation}
\label{s:2}

%
%
%
\begin{figure}
\resizebox{0.8\linewidth}{!}{
\includegraphics{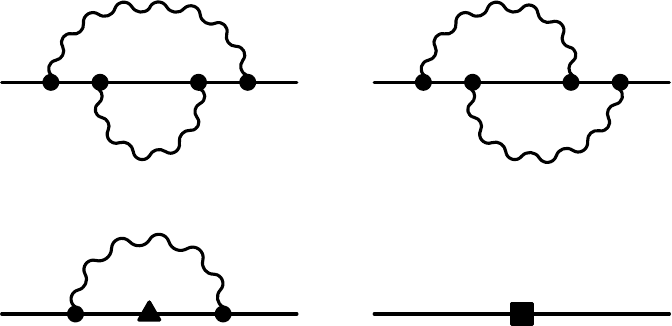}
}
\caption{
Graphical depiction of the leading 
$\cO(\alpha^2)$ 
contributions to the electromagnetic mass of charged particles in NRQED. 
The two-loop diagrams are shown in the first row, 
while the one-loop and tree-level diagrams formed using the mass counterterms are shown in the second row. 
Diagram elements are identical to those used in 
Fig.~\ref{f:oneloop}. 
Additionally the filled square denotes the leading 
$\cO(\alpha^2)$
mass counterterm. 
}
\label{f:leading2}
\end{figure}
%
%
%

Using the charge-density interaction from 
Eq.~\eqref{eq:NRQED}, 
we are led to the two, 
two-loop diagrams shown in 
Fig.~\ref{f:leading2}. 
There are additional
$\cO(\alpha^2)$
diagrams formed from the insertion of mass counterterms, 
which includes a tree-level insertion of 
$\delta_M^{(2,1)}$
and the one-loop sunset diagram with an insertion of 
$\delta_M^{(1,1)}$.

\subsubsection{Diagrams}

Evaluation of the two, two-loop diagrams is otherwise straightforward. 
We obtain an expression for the electromagnetic self energy having the form
\begin{equation}
\Sigma^{(2,1)}_\text{2-loop} 
(p_0)
= 
- (Qe)^4 
\sumintq
\sumintk
\frac{
\Sigma_{2a}(q,k,p) + \Sigma_{2b}(q,k,p)
}
{(q^2 + i \epsilon)(k^2 + i \epsilon)}
\label{eq:two}
,\end{equation}
where the momentum-dependent contribution from the double sunset diagram is given by 
\begin{eqnarray}
\Sigma_{2a}
&=&
\frac{1}{(q_0 + p_0 + i \epsilon)^2 (q_0 + k_0 + p_0 + i \epsilon)}
\label{eq:SigA}
,\end{eqnarray}
and the momentum-dependent contribution arising from the two interlocking sunsets is given by
\begin{eqnarray}
\Sigma_{2b}
&=&
\frac{1}{(q_0 + p_0 + i \epsilon) (q_0 + k_0 + p_0 + i \epsilon)(k_0 + p_0 + i \epsilon)}
.
\notag\\
\label{eq:SigB}
\end{eqnarray}

To evaluate the self energy, 
note the expectation to encounter divergent double mode sums, 
for which general methods have been developed, 
see
Refs.~\cite{Tan:2007bg,Niedermayer:2016ilf,Niedermayer:2016yll}.  
We note, 
however,  
a valuable simplification due to the static propagators appearing in the NRQED expressions. 
With the chosen momentum routing, 
the sum of the two diagrams in 
Eq.~\eqref{eq:two} 
evaluated at 
$p_0 = 0$
is proportional to the quantity
\begin{eqnarray}
\frac{1}{q_0 + k_0 + i \epsilon}
\left(
\frac{1}{q_0 + i \epsilon} + \frac{1}{k_0 + i \epsilon}
\right)
=
\frac{1}{(q_0 + i \epsilon)(k_0 + i \epsilon)}
.\notag \\
\label{eq:identity}
\end{eqnarray}
Application of this identity shows that the required double momentum mode sum 
simply factorizes into the product of two mode sums.%
\footnote{
We attribute this simplification to the non-relativistic limit employed in NRQED, 
because the analogous QED diagrams for point-like particles, 
which have the same topologies shown in Fig.~\ref{f:leading2}, 
do not seem to exhibit such simplifications. 
Such diagrams, 
moreover, 
include all finite-volume terms in the NRQED expansion, 
as well as the exponential volume dependence. 
} 
Carrying out the 
$k_0$
and
$q_0$
integrals by the residue theorem, 
we arrive at the contribution to the self energy
\begin{equation}
\Sigma^{(2,1)}_\text{2-loop} 
=
- \frac{ (Q e)^4}{4}
\,
\mathfrak{C}_3
\,
\mathfrak{C}_2
\label{eq:factorize}
.\end{equation}

Beyond 
$\cO(\alpha^2)$, 
we require the first derivative of the self-energy function evaluated at 
$p_0 = 0$. 
This derivative exhibits the same factorization property of the double mode sums, 
which requires a symmetrization of the dummy momentum modes to exhibit. 
Schematically,
there is a contribution having the form
\begin{equation}
\frac{f(q,k)}{q_0 + i \epsilon}
\longrightarrow
\frac{f(q,k)}{2}
\left(
\frac{1}{q_0 + i \epsilon}
+
\frac{1}{k_0 + i \epsilon}
\right)
\label{eq:symidentity}
,\end{equation}
where 
$f(q,k)$
is a symmetric function the under interchange of its arguments, 
$f(q,k) = f(k,q)$. 
The replacement after the arrow is achieved by renaming the dummy indices on the mode sums and integration variables. 
This allows one to utilize the identity in Eq.~\eqref{eq:identity} to cancel the non-factorized denominator. 
Carrying out this procedure, 
we determine
\begin{equation}
\Sigma_\text{2-loop}^{\prime(2,1)}
=
- 
\frac{(Qe)^4}{8} 
\left[ 
(\mathfrak{C}_3)^2
+
4 \,
\mathfrak{C}_2 \, \mathfrak{C}_4  
\right]
.\end{equation}

The final loop contribution from the figure is the one-loop diagram with leading-order counterterm insertion. 
Evaluating this diagram leads to the results
\begin{eqnarray}
\Sigma_\text{1-loop}^{(2,1)} 
&=&
- 
\delta_M^{(1,1)}
\frac{(Qe)^2}{2} 
\mathfrak{C}_3
,\notag \\
\Sigma_\text{1-loop}^{\prime (2,1)}
&=&
-
\delta_M^{(1,1)}
(Q e)^2 
\mathfrak{C}_4
.\end{eqnarray}

\subsubsection{$\cO(\alpha^2)$ Result}

Taking into account all diagrams in Fig.~\ref{f:leading2},
the leading 
$\cO(\a^2)$
self-energy can thus be written as 
\begin{eqnarray}
\Sigma^{(2,1)} 
&=& 
\Sigma^{(2,1)}_\text{2-loop} + \Sigma^{(2,1)}_\text{1-loop} + \delta_M^{(2,1)}
\notag \\
&=&
- \frac{(Qe)^2}{2} 
\left( \Delta M^{(1,0)} + \frac{(Qe)^2}{2} C_2 \right)
\mathfrak{C}_3
+ 
\delta_M^{(2,1)}
.\qquad \end{eqnarray}
From 
Eq.~\eqref{eq:Mexp_1_2}, 
the mass to the same order is given by 
\begin{equation}
M^{(2,1)}
= 
\Sigma^{(2,1)} 
+ 
M^{(1,1)} \Sigma^{\prime (1,1)}
.\end{equation}
As a consequence, 
all terms proportional to 
$\mathfrak{C}_3$
cancel, 
leaving only the counterterm contribution to the mass
$M^{(2,1)}$. 
If this were not the case, 
then the computation would have to be renormalized by subtracting the ultraviolet divergence in 
$\mathfrak{C}_3$ 
that appears in the infinite-volume limit. 
This subtraction, 
however, 
would subsequently introduce the infrared divergent integral
$\hat{\mathfrak{C}}_3$. 
The resolution of this conundrum is that the coefficient of 
$\mathfrak{C}_3$
simply vanishes. 
Thus we have no finite-volume correction at this order, 
and
\begin{equation}
M^{(2,1)} 
= 
\Delta M^{(2,0)}
,\end{equation}
where 
$\Delta M^{(2,0)}$
is the 
$\cO(\alpha^2)$
electromagnetic mass in infinite volume. 
Furthermore, 
this necessitates choosing the (finite) counterterm
\begin{equation}
\delta_M^{(2,1)}
= 
\Delta M^{(2,0)}
.\end{equation}

\subsection{Leading Three-Loop Calculation}

\begin{figure}
\resizebox{0.8\linewidth}{!}{
\includegraphics{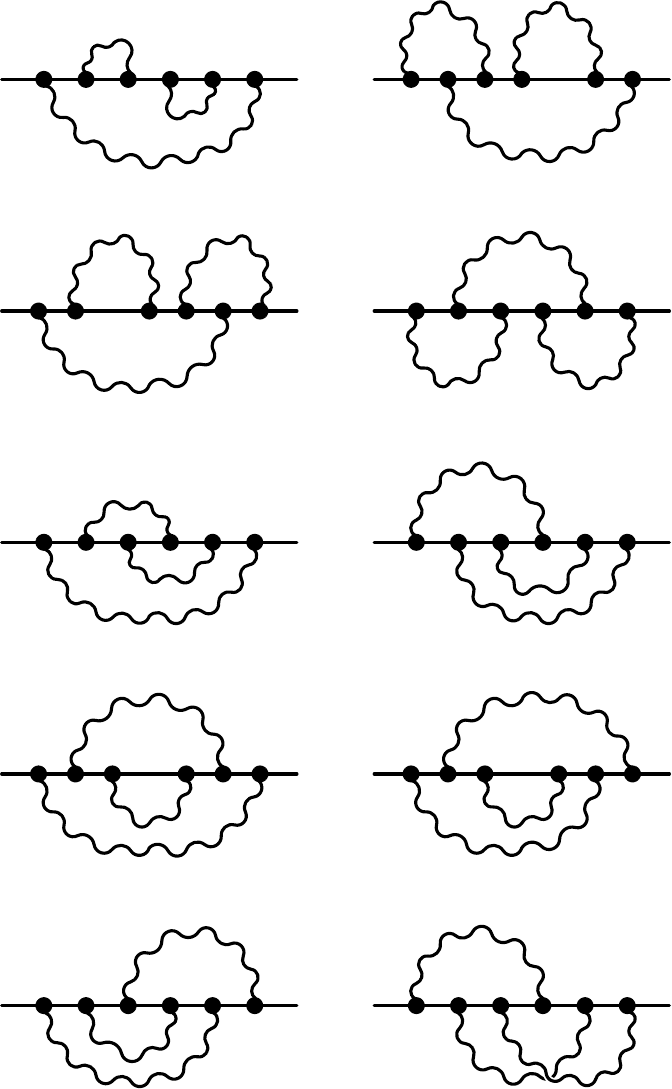}
}
\caption{
Graphical depiction of the leading three-loop diagrams contributing to 
$\Sigma^{(3,1)}$ 
in finite-volume NRQED.
Diagram elements are identical to those employed in Fig.~\ref{f:oneloop}. 
Notice that the three-loop diagram appearing on the right in the bottom row is non-planar. 
}
\label{f:Nleading3}
\end{figure}

\begin{figure}
\resizebox{0.8\linewidth}{!}{
\includegraphics{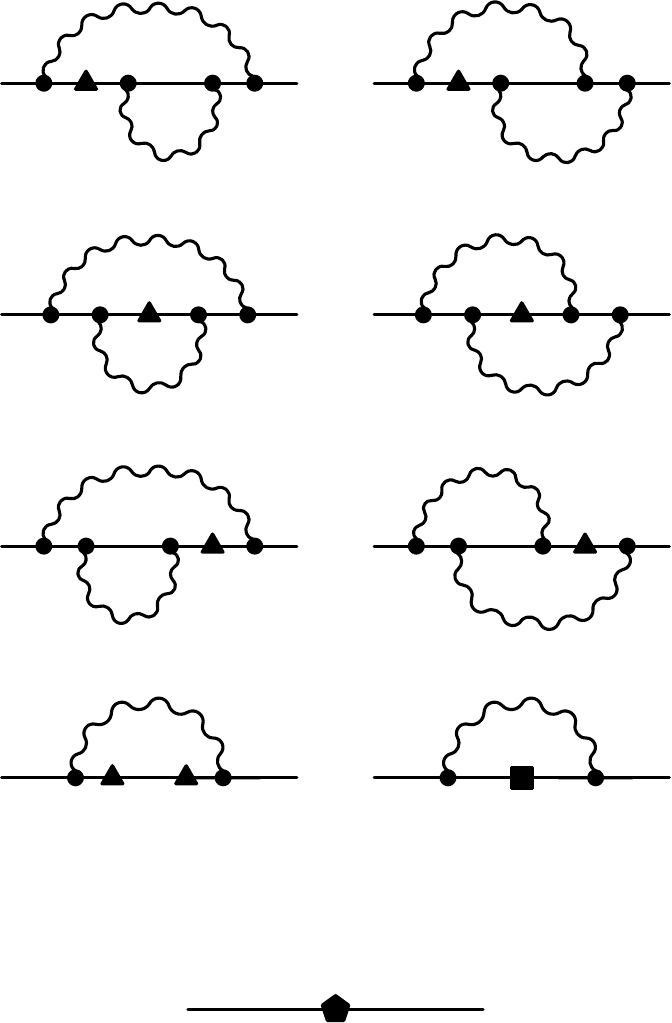}
}
\caption{
Depiction of the remaining 
$\cO(\alpha^3)$
diagrams required to determine the self-energy, 
$\Sigma^{(3,1)}$. 
These diagrams feature insertion of mass counterterms, 
where the leading mass counterterm of order 
$\alpha^n$
is shown as a filled $(n +2)$--gon. 
Photons, matter fields and charge-density vertices are as described in Fig.~\ref{f:oneloop}.  
}
\label{f:Nleading3ctms}
\end{figure}

The exact cancelation occurring at 
$\cO(\alpha^2)$ 
and the curious factorization property of the two-loop self energy 
lead us to further consider the leading three-loop calculation of the electromagnetic mass. 
The one-particle irreducible three-loop diagrams required for this calculation are shown in 
Fig.~\ref{f:Nleading3}. 
Additional diagrams formed from insertions of the mass counterterms are needed, and these appear in 
Fig.~\ref{f:Nleading3ctms}.

%
%
%
%
%
%

\subsubsection{Three-Loop Diagrams}

We begin by evaluating the three-loop diagrams.  
Note that to determine the mass to $\cO(\alpha^3)$, 
we only require the self energy evaluated at 
$p_0 = 0$. 
While the result of the calculation should be a divergent triple sum over momentum modes, 
we find that the net result of all ten diagrams can be expressed in terms of the factorized product of three, single mode sums. 
To establish this, 
we write the sum of the three-loop diagrams in the form
\begin{eqnarray}
\Sigma_\text{3-loop}^{(3,1)}
&=&
\, \sumintq
\sumintk
\sumintp
\frac{i (Qe)^6}{(q^2 + i \epsilon)(k^2 + i \epsilon)(p^2 + i \epsilon)}
\notag \\
&& \phantom{space}
\times \frac{\sum_{\beta = a, \cdots, j} \, \Sigma_{3\beta} (q_0,k_0,p_0)}{(q_0 + i \epsilon) (q_0 + k_0 + i \epsilon)}
\label{eq:3loop}
,\end{eqnarray}
where the index 
$\beta$ 
runs  alphabetically over the diagrams in 
Fig.~\ref{f:Nleading3}
(left to right, top to bottom).
We choose the momentum routing so that the first photon to interact with the matter field injects the momentum 
$q$, 
while the second photon injects 
$k$, 
and the third injects 
$p$. 
For this reason, 
all diagrams contain the three photon propagators, and the two matter field propagators appearing in Eq.~\eqref{eq:3loop}, 
which have been accordingly factored out of each 
of the residual loop functions,
$\Sigma_{3\beta} (q_0,k_0,p_0)$.  
These loop functions are merely the product of the remaining three matter-field propagators, 
and thus can all be expressed in terms of 
\begin{equation}
\xi (A, B, C) = \frac{1}{(A + i \epsilon)(B + i \epsilon)(C+ i \epsilon)}
.\end{equation}
Evaluation of the diagrams produces
\begin{eqnarray}
\Sigma_{3a} 
&=&
\xi (q_0, q_0 + p_0, q_0), 
\notag \\
\Sigma_{3b} 
&=&
\xi(q_0, q_0 + p_0, p_0)
,\notag \\
\Sigma_{3c}
&=&
\xi(k_0,k_0+p_0,k_0)
,\notag\\
\Sigma_{3d}
&=&
\xi(k_0,k_0+p_0,p_0)
,\notag\\
\Sigma_{3e}
&=&
\xi(q_0+k_0+p_0,q_0+p_0,q_0)
,\notag \\
\Sigma_{3f}
&=&
\xi(q_0+k_0+p_0,q_0+p_0,p_0)
,\notag\\
\Sigma_{3g}
&=&
\xi(q_0+k_0+p_0,q_0+k_0,q_0)
,\notag\\
\Sigma_{3h}
&=&
\xi(q_0+k_0+p_0,q_0+k_0,k_0)
,\notag\\
\Sigma_{3i}
&=&
\xi(q_0+k_0+p_0,k_0+p_0,k_0)
,\notag\\
\Sigma_{3j}
&=&
\xi(q_0 + k_0 + p_0,k_0+p_0,p_0)
,\end{eqnarray}
where we have kept the ordering of the momentum arguments 
(left to right) 
to correspond to the progression in time 
(right to left)
in the diagrams.

Alphabetical ordering of the diagrams has been chosen so that, 
when summed in pairs, 
the identity in Eq.~\eqref{eq:identity}
can be utilized. Applying this identity to the  five pairs, 
we have
\begin{eqnarray}
\Sigma_{3a} 
+ 
\Sigma_{3b}
&=&
\xi(q_0,q_0,p_0)
,\notag \\
\Sigma_{3c} 
+ 
\Sigma_{3d}
&=&
\xi(k_0, k_0, p_0)
,\notag \\
\Sigma_{3e} 
+ 
\Sigma_{3f}
&=&
\xi(q_0+ k_0 + p_0, q_0, p_0)
,\notag \\
\Sigma_{3g} 
+ 
\Sigma_{3h}
&=&
\xi(q_0+ k_0 + p_0, q_0, k_0)
,\notag \\
\Sigma_{3i} 
+ 
\Sigma_{3j}
&=&
\xi(q_0+ k_0 + p_0, k_0, p_0)
.\end{eqnarray}
On account of the further identity 
\begin{eqnarray}
\xi(A,B,C)
&=&
\xi(A+B+C,A,B) 
+ 
\xi(A+B+C,A,C) 
\notag \\
&& \phantom{spacing}
+ \xi(A+B+C,B,C) 
,\end{eqnarray}
we obtain a simple result for the sum of these ten contributions, 
namely
\begin{equation}
\sum_{\beta= a, \cdots, j} 
\, 
\Sigma_{3 \beta}
=
\xi(q_0,q_0,p_0)
+
\xi(k_0, k_0, p_0)
+
\xi(q_0,k_0,p_0)
.\end{equation}
Consequently the triple mode sum factorizes into a product of a single mode sum
(over the momentum labelled by $p$), 
and a double mode sum.
Notice that the latter is not factorized at this stage due to the propagator 
$(q_0 + k_0 + i \epsilon)^{-1}$
appearing in Eq.~\eqref{eq:3loop}.

Reorganizing the terms appearing in the three-loop contribution, 
Eq.~\eqref{eq:3loop}, 
to account for the dramatic simplifications uncovered so far, 
we reduce the expression for the self energy to
\begin{eqnarray}
\Sigma_\text{3-loop}^{(3,1)}
&=&
- \frac{(Qe)^6}{2} \mathfrak{C}_2
\,
\sumintq
\sumintk
\frac{\sigma(q_0,k_0)}{(q^2 + i \epsilon)(k^2 + i \epsilon)}
\label{eq:3loop2nd}
,\end{eqnarray}
where the function 
$\sigma(q_0,k_0)$
is a product of four matter-field propagators, 
and has the form
\begin{eqnarray}
\sigma
&=&
(q_0 + i \epsilon)^{-1}
\Big[
\xi(q_0+k_0, q_0, q_0)
+ 
\xi(q_0+k_0, q_0, k_0)
\notag \\
&& \phantom{spacingses}
+
\xi(q_0+k_0, k_0, k_0)
\Big]
.\end{eqnarray}
Finally, 
we symmetrize with respect to the momentum labels, 
which results in the replacement
\begin{eqnarray}
\sigma(q_0, k_0) 
&\longrightarrow&
\frac{1}{2} \left[ \sigma(q_0, k_0) + \sigma(k_0, q_0) \right]
\notag \\
&=&
\frac{\xi(q_0, q_0, k_0)
+ 
\xi(q_0, k_0, k_0)
+
\xi(k_0, k_0, k_0)}{2 (q_0 + i \epsilon)}
,\notag \\
\end{eqnarray}
where the second line results upon application of the identity in 
Eq.~\eqref{eq:identity}. 
In this form, 
we have decoupled the remaining two momentum mode sums, 
which can then be expressed in terms of the 
$\mathfrak{C}_j$
defined in Eq.~\eqref{eq:Cj}.  
The final result is
\begin{eqnarray}
\Sigma_\text{3-loop}^{(3,1)}
=
\frac{(Qe)^6}{16} \mathfrak{C}_2
\left[
\left( \mathfrak{C}_3 \right)^2
+ 
2 \, 
\mathfrak{C}_2 \, \mathfrak{C}_4
\right]
.\end{eqnarray}

%
%
%
%
%
%

\subsubsection{Remaining Diagrams}

The remaining contributions to the self-energy at 
$\cO(\alpha^3)$
are shown in 
Fig.~\ref{f:Nleading3ctms}.
These are: 
the two-loop diagrams with an insertion of the $\cO(\a)$ counterterm, 
the one-loop diagram with two insertions of the $\cO(\a)$ counterterm, 
the one-loop diagram with an insertion of the $\cO(\a^2)$ counterterm, 
and finally the tree-level diagram arising from the 
$\cO(\a^3)$
counterterm.

Evaluation of the two-loop diagrams with an insertion of 
$\delta_{M}^{(1,1)}$
is similar to the calculation of derivative of the two-loop self energy. 
Not surprisingly, 
it can be expressed in terms of that quantity. 
We find
\begin{equation}
\Sigma_\text{2-loop}^{(3,1)}
=
-
\delta_M^{(1,1)}
\Sigma_\text{2-loop}^{\prime (2,1)}
\label{eq:twoloopMins}
,\end{equation}
where the negative sign can easily be explained as from the difference between inserting a mass and requiring an additional propagator, 
$\frac{i}{q_0 + i \epsilon} ( - i \delta_M) \frac{i}{q_0 + i \epsilon}$, 
versus
differentiation of a propagator with respect to the external momentum, 
$\frac{d}{dp_0} \frac{i}{q_0 + p_0 + i \epsilon} \big|_{p_0=0} =  - \frac{i}{(q_0 + i \epsilon)^2}$.

The one-loop diagrams are also readily evaluated. 
The diagram with a single insertion of a mass counterterm has already been determined above, 
and can be expressed in terms of (minus) the derivative of the one-loop self energy. 
The diagram with two insertions of the leading-order mass counterterm can be expressed in terms of the second derivative of the one-loop
self energy. 
For this contribution, 
there is no sign difference due to two differentiations, 
however, 
there is a factor of $2$ difference compared with differentiation. 
Hence, 
we find
\begin{equation}
\Sigma_\text{1-loop}^{(3,1)} 
=
- 
\delta_M^{(2,1)} \Sigma_\text{1-loop}^{\prime(1,1)}
+
\frac{1}{2} \left( \delta_M^{(1,1)} \right)^2  \Sigma_\text{1-loop}^{\prime \prime (1,1)}
.\end{equation}
Finally there is the local contribution from the counterterm 
$\delta_M^{(3,1)}$.

\subsubsection{$\cO(\alpha^3)$ Result}

%
\begin{table}
\caption{%
Catalogue of results obtained for contributions to the self energies, the mass counterterms, and the electromagnetic mass. 
Notice the 
$\cO(\alpha^2)$ 
results are all expressible as factorized products of the 
$\cO(\alpha)$
results. 
}
\begin{center}
\resizebox{0.8\linewidth}{!}{
\begin{tabular}{|c|c|}
\hline
\hline
\multicolumn{2}{|c|}{$\cO(\alpha)$}
\tabularnewline
\hline
\hline
$\quad \Sigma_\text{1-loop}^{(1,1)} \quad$ 
&
$\frac{1}{2} (Qe)^2 \mathfrak{C}_2$
\tabularnewline
$\Sigma_\text{1-loop}^{\prime(1,1)}$ 
&
$\frac{1}{2} (Qe)^2 \mathfrak{C}_3$
\tabularnewline
$\Sigma_\text{1-loop}^{\prime\prime(1,1)}$ 
&
$(Qe)^2  \mathfrak{C}_4$
\tabularnewline 
$\delta_M^{(1,1)}$
&
$\Delta M^{(1,0)} - \frac{1}{2} (Qe)^2 \hat{\mathfrak{C}}_2$
\tabularnewline
$M^{(1,1)}$
&
$\Delta M^{(1,0)} + \frac{1}{2} (Qe)^2 C_2$
\tabularnewline
[0.5mm]
\hline
\hline
\multicolumn{2}{c}{}
\tabularnewline
\hline
\hline
\multicolumn{2}{|c|}{$\cO(\alpha / M)$}
\tabularnewline
\hline
\hline
$\Sigma_\text{1-loop}^{(1,2)}$ 
&
$\frac{1}{2M} (Qe)^2 \mathfrak{C}_1$
\tabularnewline
$\Sigma_\text{1-loop}^{\prime(1,2)}$ 
&
$-\frac{1}{2M} (Qe)^2 \mathfrak{C}_2$
\tabularnewline
$\delta_M^{(1,2)}$
&
$ - \frac{1}{2M} (Qe)^2 \hat{\mathfrak{C}}_1$
\tabularnewline
$M^{(1,2)}$
&
$ \frac{1}{2M} (Qe)^2 C_1$
\tabularnewline
[0.5mm]
\hline
\hline
\multicolumn{2}{c}{}
\tabularnewline
\hline
\hline
\multicolumn{2}{|c|}{$\cO(\alpha^2)$}
\tabularnewline
\hline
\hline
$\quad \Sigma_\text{2-loop}^{(2,1)} \quad$ 
&
$- \Sigma_\text{1-loop}^{(1,1)} \Sigma_\text{1-loop}^{\prime(1,1)}$
\tabularnewline
[0.5mm]
$\Sigma_\text{2-loop}^{\prime(2,1)}$ 
&
$-\Sigma_\text{1-loop}^{(1,1)} \Sigma_\text{1-loop}^{\prime \prime(1,1)}  - \frac{1}{2} (\Sigma_\text{1-loop}^{\prime(1,1)})^2$
\tabularnewline
[0.5mm]
$\Sigma_\text{1-loop}^{(2,1)}$ 
&
$- \delta_M^{(1,1)} \Sigma_\text{1-loop}^{\prime(1,1)}$
\tabularnewline
[0.5mm]
$\Sigma_\text{1-loop}^{\prime(2,1)}$ 
&
$- \delta_M^{(1,1)} \Sigma_\text{1-loop}^{\prime\prime(1,1)}$
\tabularnewline 
$\delta_M^{(2,1)}$
&
$\Delta M^{(2,0)}$
\tabularnewline
$M^{(2,1)}$
&
$\Delta M^{(2,0)}$
\tabularnewline
\hline
\hline
\end{tabular}
}
\end{center}
\label{t:summy}
\end{table}

All of the contributions are determined for the electromagnetic mass at leading three-loop order. 
The self energy takes the form
\begin{eqnarray}
\Sigma^{(3,1)}
&=&
\Sigma_\text{3-loop}^{(3,1)}
+ 
\Sigma_\text{2-loop}^{(3,1)}
+ 
\Sigma_\text{1-loop}^{(3,1)}
+ 
\delta_M^{(3,1)}
.\end{eqnarray} 
To obtain the mass, 
we utilize the leading-order form of 
Eq.~\eqref{eq:Mexp_3}, 
which is
\begin{eqnarray}
M^{(3,1)} 
&=& 
\Sigma^{(3,1)}
+ 
M^{(2,1)} \Sigma^{\prime(1,1)}
\notag \\
&& 
+ 
M^{(1,1)} \Sigma^{\prime(2,1)}
+ 
\frac{1}{2}  \left( M^{(1,1)} \right)^2   \Sigma^{\prime \prime (1,1)}  
,\, \,
\label{eq:Mexp_3_1}
\end{eqnarray}
where the required $\cO(\alpha)$ 
results are
$\Sigma^{\prime (1,1)} = \Sigma^{\prime (1,1)}_\text{1-loop}$
and
$\Sigma^{\prime \prime (1,1)} = \Sigma^{\prime \prime (1,1)}_\text{1-loop}$. 
The required
$\cO(\alpha^2)$
result is
$\Sigma^{\prime(2,1)} = \Sigma_\text{2-loop}^{\prime(2,1)} + \Sigma_\text{1-loop}^{\prime (2,1)}$.
For convenience, 
a summary of these results appears in 
Table~\ref{t:summy}. 
Combining all contributions to the electromagnetic mass, 
there are numerous cancelations, 
leaving us with the reduced expression
\begin{eqnarray}
M^{(3,1)}
&=&
\Sigma_\text{3-loop}^{(3,1)}
+ 
\Sigma_\text{1-loop}^{(1,1)} 
\Sigma_\text{2-loop}^{\prime(2,1)}
\notag \\
&& \phantom{spa}
+
\frac{1}{2} 
\left(\Sigma_\text{1-loop}^{(1,1)} \right)^2
\Sigma_\text{1-loop}^{\prime \prime (1,1)} 
+
\delta_M^{(3,1)}
\label{eq:reduced}
.\end{eqnarray}
Utilizing the results from the table, 
we find the simple result
\begin{eqnarray}
M^{(3,1)}
=
\delta_M^{(3,1)}
,\end{eqnarray}
where the value of the counterterm must be
\begin{equation}
\delta_M^{(3,1)} 
= 
\Delta M^{(3,0)}
,\end{equation}
which is the 
$O(\alpha^3)$ 
electromagnetic mass in infinite volume. 
Indeed the non-renormalizable 
$\left(\mathfrak{C}_3\right)^2$
contributions to the electromagnetic mass necessarily cancel. 
Other contributions at this order are proportional to 
$(\mathfrak{C}_2)^2 \mathfrak{C}_4$, 
and these too cannot be renormalized by subtracting the infinite-volume limit; 
because, 
in the case of 
$\mathfrak{C}_4$, 
this subtraction introduces an infrared divergence.%
\footnote{
Whereas 
$\mathfrak{C}_3$
is divergent in the ultraviolet, 
the sums
$\mathfrak{C}_{j>3}$
are all finite, but do not possess an infinite-volume limit. 
}
In fact, 
the vanishing of the finite-volume effect can be seen from the simplified expression in 
Eq.~\eqref{eq:reduced}. 
Absent from this expression are lower-order counterterms, 
which implies that all contributions are products of three unrenormalized mode sums. 
Because such products cannot be renormalized by a single counterterm, 
the sum of loop corrections vanishes. 
Thus there is no finite-volume effect at leading three-loop order.

\subsection{Next-To-Leading Two-Loop Calculation}

The leading two-loop and three-loop contributions to the electromagnetic mass exhibit a factorization property. 
As a result, the divergent double and triple mode sums can be written as the product of single mode sums 
after summing over all contributing Feynman diagrams. 
Vanishing finite-volume effects are obtained, 
moreover, 
due to the impossibility to renormalize certain single momentum mode sums appearing in the computation. 
These curious properties lead us to our final computation, 
namely the calculation of the next-to-leading two-loop electromagnetic mass.

%
%
%
\begin{figure}
\resizebox{\linewidth}{!}{
\includegraphics{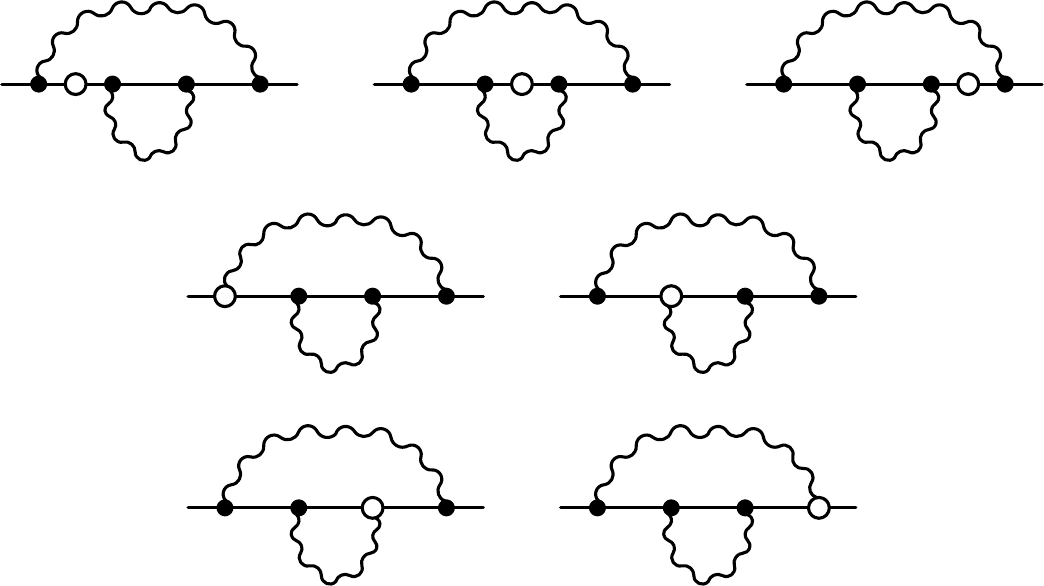}
}
\vskip 0.15in
\resizebox{\linewidth}{!}{
\includegraphics{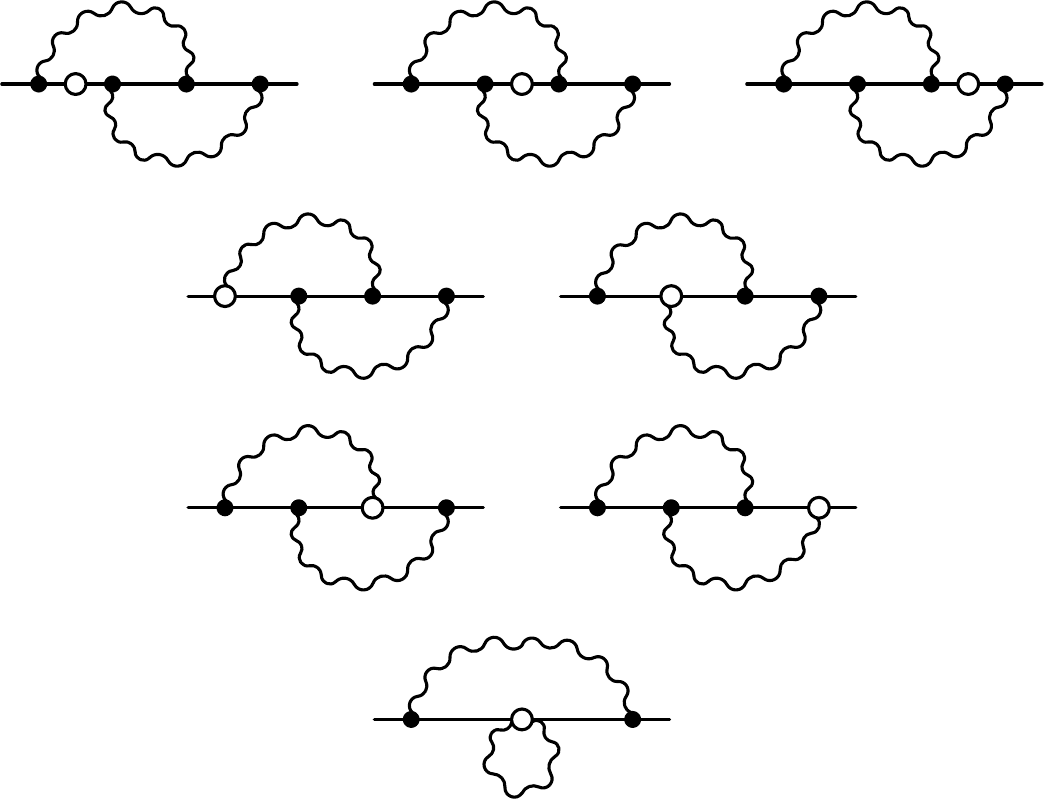}
}
\caption{
Graphical depiction of the next-to-leading order two-loop diagrams in finite-volume NRQED.
Diagram elements are the same as those appearing in Fig.~\ref{f:oneloop}. 
}
\label{f:Nleading2}
\end{figure}
%
%
%

At next-to-leading order, 
one has an additional operator in the NRQED Lagrangian density, 
which is the gauge-covariant kinetic energy, 
Eq.~\eqref{eq:KinE}. 
This operator introduces the hadron's mass $M$, 
but results otherwise remain spin and structure independent. 
Three new interaction vertices, which differ by the number of photons, 
are generated from the kinetic term. 
All possible two-loop diagrams formed from the insertion of one of these next-to-leading order vertices are shown in 
Fig.~\ref{f:Nleading2}.
There are additional one-loop diagrams contributing at this order formed from the insertion of lower-order counterterms.
These are depicted in 
Fig.~\ref{f:Nleading1}

\subsubsection{Next-To-Leading Two-Loop Diagrams}

There are three basic topologies encountered at next-to-leading two-loop order:
(a)~double sunset, 
(b)~interlocking sunsets, 
and
(c)~tadpole riding into the sunset.
These are shown in the first through third rows, 
the fourth through sixth rows, 
and the seventh row of 
Fig.~\ref{f:Nleading2}, 
respectively. 
Accordingly, 
we write the regulated self energy as a sum of the three corresponding contributions
\begin{equation}
\Sigma_\text{2-loop}^{(2,2)}
= 
- \frac{(Qe)^4}{2M}
\left[
\Sigma^{(2,2)}_{2a} + \Sigma^{(2,2)}_{2b} + \Sigma^{(2,2)}_{2c}
\right].
\label{eq:sumNLO}
\end{equation} 
The overall multiplicative factor is chosen for convenience.

As many of the diagrams are related by gauge invariance, 
we choose to economize the computation by adopting Feynman gauge. 
In this gauge, 
the photon propagator is diagonal in that temporal and spatial components of the photon field do not mix. 
Consequently, 
all diagrams in the second, third, fifth and sixth rows of Fig.~\ref{f:Nleading2} vanish. 
To evaluate diagrams with the double sunset topology
(a), 
the non-vanishing diagrams appear in the first row.
Such diagrams 
feature the insertion of the kinetic-energy operator,  
and an additional matter-field propagator compared to the leading two-loop order.
Consequently the contribution from these diagrams can be written in the form 
\begin{equation}
\Sigma_{2b}^{(2,2)} = 
\sumintq \sumintk
\Sigma_{2a} (q,k,0) 
\frac{
2 \, T(q)
+ 
T(q+k)}
{(q^2 + i \epsilon) (k^2 + i \epsilon)}
,\end{equation}
where 
$\Sigma_{2a}(q,k,p)$ 
is the function of loop momenta given in 
Eq.~\eqref{eq:SigA}, 
and 
$T(q)$ 
accounts for insertion of the kinetic-energy operator with an accompanying matter-field propagator. 
Specifically, 
it is given by
\begin{eqnarray}
T(q)
= 
\frac{\bm{q}^2}{q_0 + i \epsilon}
.\end{eqnarray}
Non-vanishing diagrams with kinetic-energy insertion in the interlocking sunsets, 
which have topology (b), 
are the diagrams appearing in the fourth row of 
Fig.~\ref{f:Nleading2}.  
We can similarly write their contribution in the form
\begin{equation}
\Sigma_{2b}^{(2,2)} 
= \sumintq \sumintk
\Sigma_{2b} (q,k,0) 
\frac{
T(q)
+
T(q+k)
+
T(k)}
{(q^2 + i \epsilon)(k^2 + i \epsilon)}
,\end{equation}
where 
$\Sigma_{2b}(q,k,p)$ 
is the function of loop momenta defined in 
Eq.~\eqref{eq:SigB}.

Due to the symmetry property under momentum interchange, 
namely
$\Sigma_{2b} (k,q,p) = \Sigma_{2b}(q,k,p)$, 
the sum of contributions from (a) and (b) topologies has the form 
\begin{eqnarray}
\Sigma^{(2,2)}_{2a} + \Sigma^{(2,2)}_{2b} 
&=&
\sumintq \sumintk
\left[ \Sigma_{2a} (q,k,0) + \Sigma_{2b} (q,k,0) \right]
\notag \\
&& \phantom{spacing}
\times \frac{2 \, T(q) + T(q+k)}{(q^2 + i \epsilon)(k^2 + i \epsilon)}
\label{eq:twoNLO}
.\end{eqnarray}
Notice that because the sum of these contributions is proportional to 
$\Sigma_{2a}(q,k,0) + \Sigma_{2b}(q,k,0)$, 
the identity in Eq.~\eqref{eq:identity} can be utilized to produce
\begin{eqnarray}
\Sigma_{2a}(q,k,0) + \Sigma_{2b}(q,k,0)
=
\xi(q_0,q_0,k_0)
.\end{eqnarray}
At this stage, 
the only remaining non-factorized contribution from these diagrams involves the factor of
$T(q+k)$
appearing above.  
Due to cubic symmetry, 
however,
we can effectively replace
\begin{equation}
T(q+k) \to \frac{\bm{q}^2 + \bm{k}^2}{q_0 + k_0 + i \epsilon}
,\end{equation}
because the 
$\bm{q} \cdot \bm{k}$
piece sums to zero. 
For the 
$\bm{k}^2$
contribution from this term, 
we can interchange 
$k$ 
and 
$q$ 
momentum labels in Eq.~\eqref{eq:twoNLO}
to write it as a 
$\bm{q}^2$
contribution.
The sum of two such 
$\bm{q}^2$
contributions allows for another application of the identity appearing in Eq.~\eqref{eq:identity}, 
because we have
\begin{eqnarray}
\frac{(\bm{q}^2 + \bm{k}^2)
\xi(q_0,q_0,k_0)
}{q_0 + k_0 + i \epsilon}
&\to&
\frac{\bm{q}^2 \left[
\xi(q_0,q_0,k_0)
+ 
\xi(k_0,k_0,q_0)
\right]
}{q_0 + k_0 + i \epsilon}
\notag \\
&=&
\frac{\bm{q}^2}{(q_0 + i \epsilon)^2}
\frac{1}{(k_0 + i \epsilon)^2}
.\end{eqnarray}
Consequently the double mode sums from topologies 
(a) and (b), 
given in 
Eq.~\eqref{eq:twoNLO}, 
can be written in the factorized form 
\begin{eqnarray}
\Sigma_{2a}^{(2,2)}
+
\Sigma_{2b}^{(2,2)}
&=&
-
\frac{1}{4}
\Big[
\mathfrak{C}_1 \, 
\mathfrak{C}_3
+ 
2 
\left(\mathfrak{C}_2 \right)^2
\Big]
.
\end{eqnarray}

The tadpole subdiagram riding into the sunset trivially factorizes into a product of two momentum mode sums, 
because the sunset loop momentum does not flow into the tadpole loop. 
Straightforward computation leads to this self-energy contribution
\begin{equation}
\Sigma_{2c}^{(2,2)}
=
\frac{3}{4}
\mathfrak{C}_1 \, 
\mathfrak{C}_3
,\end{equation}
where the mode sums appear in Eq.~\eqref{eq:Cj}.
Taking the sum of all three topologies results in 
\begin{equation}
\Sigma_\text{2-loop}^{(2,2)}
=
- \frac{(Qe)^4}{4 M}
\left[
\mathfrak{C}_1 \, 
\mathfrak{C}_3
-
\left(\mathfrak{C}_2 \right)^2
\right]
.\end{equation}

\subsubsection{Additional One-Loop Diagrams}

%
%
%
\begin{figure}
\resizebox{0.8\linewidth}{!}{
\includegraphics{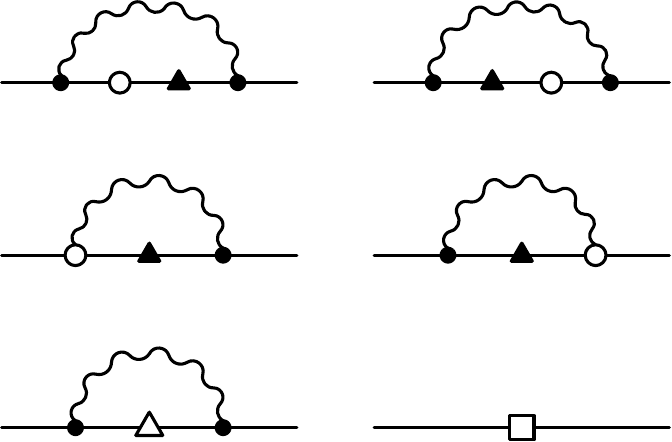}
}
\caption{
Additional diagrams required in computing the electromagnetic mass to 
$\cO(\alpha^2 / M)$. 
Diagram elements are the same as those appearing in Fig.~\ref{f:oneloop}. 
The open square denotes the counterterm, 
$\delta_M^{(2,2)}$
necessary to renormalize the computation of the mass. 
}
\label{f:Nleading1}
\end{figure}
%
%
%

At the next-to-leading two-loop order, 
one additionally requires one-loop diagrams formed from the insertion of lower-order mass counterterms, 
as shown in Fig.~\ref{f:Nleading1}, 
as well as the tree-level counterterm, 
$\delta_M^{(2,2)}$. 
Contributions from the one-loop diagrams are of two types.
The (a) type contribution arises from the insertion of the leading-order mass counterterm, 
$\delta_M^{(1,1)}$, 
into the next-to-leading order one-loop diagram. 
The (b) type contribution, on the other hand, 
arises from insertion of the next-to-leading order mass counterterm, 
$\delta_M^{(1,2)}$, 
into the leading one-loop diagram.  
To separate these contributions, 
we write 
\begin{equation}
\Sigma_\text{1-loop}^{(2,2)}
=
\Sigma_{1a}^{(2,2)}
+ 
\Sigma_{1b}^{(2,2)}
.\end{equation}

Computation of the next-to-leading order diagrams with an insertion of 
$\delta_M^{(1,1)}$
can be obtained from those at next-to-leading order without an insertion. 
The crucial realization is that the additional propagator present for the mass insertion is almost identical to taking the derivative with respect to an external momentum 
$p_0$. 
Such a derivative produces a factor of two, 
however, 
this effectively accounts for there being possible two mass-insertion diagrams, 
and only one diagram without mass insertion. 
Thus it is straightforward to verify that
\begin{equation}
\Sigma_{1a}^{(2,2)}
= 
-
\delta_M^{(1,1)}
\, \Sigma^{\prime (1,2)}_\text{1-loop}
.\end{equation}
Insertion of 
$\delta_M^{(1,2)}$
in the leading-order one-loop diagram yields the contribution
\begin{equation}
\Sigma_{1b}^{(2,2)}
=
- 
\delta_M^{(1,2)} \, \Sigma_\text{1-loop}^{\prime (1,1)}
,\end{equation}
which exhausts the diagrams required at this order.

\subsubsection{$\cO(\alpha^2 / M)$ Result}

Above we have determined the self-energy function at next-to-leading two-loop order. 
It takes the form of a sum of three terms
\begin{eqnarray}
\Sigma^{(2,2)}
=
\Sigma^{(2,2)}_\text{2-loop}
+ 
\Sigma^{(2,2)}_\text{1-loop}
+
\delta_M^{(2,2)}
.\end{eqnarray}
From Eq.~\eqref{eq:Mexp_1_2}, 
the electromagnetic mass at this order in the NRQED expansion is determined by 
\begin{eqnarray}
M^{(2,2)}
=
\Sigma^{(2,2)}
+
M^{(1,1)} \Sigma^{\prime(1,2)}
+ 
M^{(1,2)} \Sigma^{\prime(1,1)}
.\end{eqnarray}
Each of these additional contributions from wavefunction renormalization exactly cancels the factorized terms in the self energy. 
We arrive at the result
$M^{(2,2)} = \delta_M^{(2,2)} = 0$, 
and there are again no finite-volume effects at this order.

\section{Summary of Results}
\label{s:summy}

We investigate corrections to hadron electromagnetic masses for larger-than-physical values of the electric charge. 
Such results are useful for understanding the systematic uncertainties of lattice QCD + QED calculations 
that employ such an approach to combat the smallness of QED effects. 
Our focus is on the long-distance modification to hadron masses arising in finite volume, 
as these can be calculated in a model-independent fashion using the framework of NRQED applied to bound-state hadrons. 
Using the actual size of unphysical values of the electric charge employed recently, 
we argue for the necessity of computing higher-order QED volume corrections in 
Sec.~\ref{s:0}. 
We compute the leading two-loop, 
leading three-loop, 
and next-to-leading two-loop electromagnetic masses in NRQED. 
These computations, 
which are presented in 
Sec.~\ref{s:HOT}, 
are considerably technical, 
however, 
general simplifying features are encountered.
The multi-loop diagrams exhibit a regularization-independent momentum factorization, 
whereby complicated double and triple momentum mode sums can be written as products of single mode sums. 
This factorization property, 
moreover, 
leads to rather dramatic cancelations between self-energy contributions and those from wavefunction renormalization. 
Consequently,
we find all three higher-order finite-volume corrections vanish. 
Non-trivial checks of the calculations are performed in the Appendices. 
In Appendix~\ref{s:A}, 
the Ward identity is verified in bare perturbation theory up to next-to-leading one-loop order, 
and leading two-loop order. 
This confirms part of our next-to-leading two-loop and leading three-loop computations, respectively. 
Additionally in Appendix~\ref{s:B}, 
results at the leading two-loop order are alternately obtained using renormalized perturbation theory. 
These results are also confirmed directly from bare perturbation theory. 
Based on the vanishing power-law effects determined here, 
we conclude that higher-order QED finite-volume corrections cannot appreciably affect computations for larger-than-physical electric charges. 
Such corrections could occur beyond the order considered in this work, 
$\sim \alpha^4 / L$ 
in the QED expansion
and 
$\sim \alpha^2 / (M^2 L^3)$
in the NRQED expansion, 
or from exponential dependence on the volume.

\begin{acknowledgments}

This work was supported in part by a grant from the Professional Staff Congress of The CUNY, 
and by the U.S. National Science Foundation, under Grant No. PHY15-15738.
BCT was additionally supported by a joint 
The City College of New York-RIKEN/Brookhaven Research Center fellowship.
BCT thanks the Institute for Nuclear Theory
for partial support during the intermediate stages of this work, 
and the organizers of 
``INT-16-1: Nuclear Physics from Lattice QCD"
for providing a stimulating environment. 
BCT also thanks the Kavli Institute for Theoretical Physics for their hospitality during the final stages of this work, 
and partial support from the U.S. National Science Foundation, under Grant No. PHY11-25915.

\end{acknowledgments}

\appendix

\section{Ward Identity}
\label{s:A}

In the main text, 
we compute the pole position of the two-point correlation function in a combined QED and NRQED expansion. 
Here we verify the non-renormalization of charge by computing the charge vertex 
up to next-to-leading one-loop order and leading two-loop order. 
The wavefunction renormalization at these respective orders has been utilized in the computation of the electromagnetic mass, 
and serves as a useful check of bare perturbation theory.

Analysis of the two-point correlation function in Eq.~\eqref{eq:2pt} leads to the wavefunction renormalization factor
\begin{equation}
Z^{-1} = 1 - \frac{d \Sigma (p_0)}{d p_0} \Big|_{p_0 = \cM} 
,\end{equation}
which is the residue at the pole,
$p_0 = \cM$. 
Expanding this factor in QED perturbation theory, 
we have 
$Z^{(0)}=1$, 
along with
\begin{eqnarray}
Z^{(1)}
&=&
\Sigma^{\prime (1)}
,\notag \\
Z^{(2)}
&=&
\Sigma^{\prime (2)}
+
\left( \Sigma^{\prime (1)} \right)^2
+
M^{(1)} \Sigma^{\prime\prime (1)}
\label{eq:Zs}
,\end{eqnarray}
where the latter requires the first term in the perturbative expansion of the pole about zero. 
For the former, 
we shall also expand to leading and next-to-leading order in NRQED, 
to arrive at
$Z^{(1,1)} = \Sigma^{\prime (1,1)}$
and
$Z^{(1,2)} = \Sigma^{\prime (1,2)}$.

%
%
%
\begin{figure}
\resizebox{0.7\linewidth}{!}{
\includegraphics{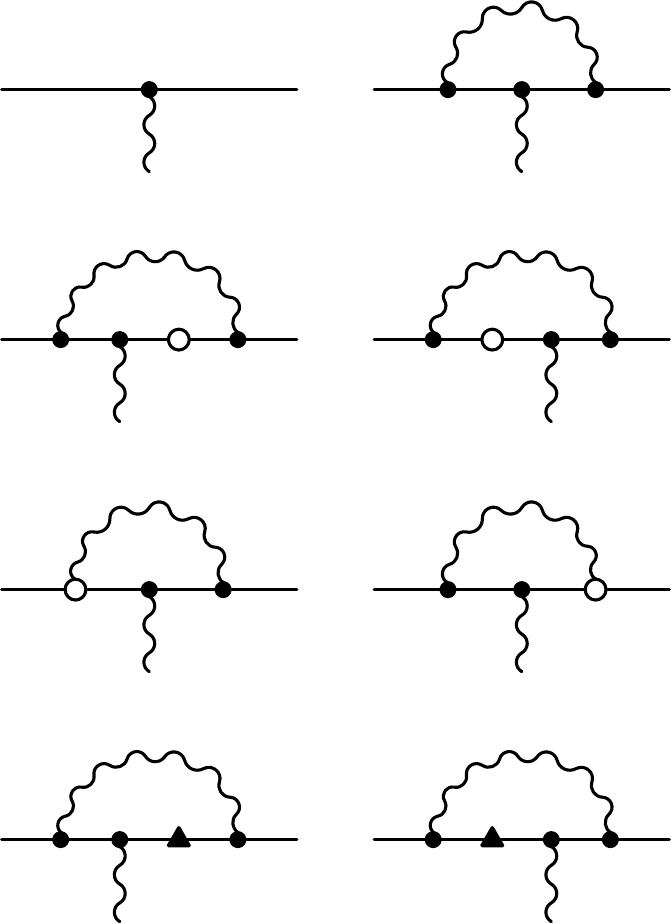}
}
\caption{
Graphical depiction of contributions to the three-point function of the 
charge-density interaction. 
The top row shows the tree-level vertex, 
$\Gamma^{(0)}$, 
and leading one-loop correction, 
$\Gamma^{(1,1)}$. 
The middle two rows show the next-to-leading one-loop corrections, 
$\Gamma^{(1,2)}$.
Finally, 
the last row depicts the one-loop contributions to 
$\Gamma^{(2,1)}$, 
which feature an insertion of the leading-order mass counterterm, 
$\delta_M^{(1,1)}$. 
Diagram elements are the same as those appearing in 
Fig.~\ref{f:oneloop}. 
}
\label{f:VLONLO}
\end{figure}
%
%
%

The reducible three-point function involving the charge-density interaction,
$\rho = Q e \, \psi^\dagger \psi$, 
is denoted
$\Gamma_\text{reducible}(p'_0,p_0)$, 
and is given by 
\begin{equation}
\Gamma_\text{reducible} (p'_0,p_0) 
= 
G(p'_0)  \, 
\Gamma (p'_0,p_0) \, 
G(p_0)
,\end{equation}
where the one-particle irreducible contribution, 
$\Gamma (p'_0,p_0)$, 
has been separated. 
The amputated three-point function is of interest to us, 
and is conveniently written in terms of a vertex function, 
$V$,  
as 
$\Gamma_\text{amputated} \equiv Qe \, V$. 
The amputated three-point function is obtained by taking the normalized residues at the poles. 
In momentum space, 
this is achieved by dividing by the two-point functions
\begin{eqnarray}
Q e \, 
V
&=& 
\lim_{p'_0, p_0 \to \cM} 
\frac{\Gamma_\text{reducible} (p'_0, p_0)}{Z^{-\frac{1}{2}} G(p'_0) Z^{-\frac{1}{2}} G(p_0)}
=
Z \, \Gamma (\cM, \cM)
.\notag \\
\end{eqnarray}
The one-particle irreducible three-point function has the QED expansion to second order
\begin{equation}
\Gamma (\cM, \cM)
=
\Gamma^{(0)}
+
\Gamma^{(1)}
+ 
\Gamma^{(2)}
+ 
M^{(1)} \, 
\Gamma^{\prime (1)}
,\end{equation}
where all suppressed arguments are implicitly evaluation at
$p_0 = 0$. 
Combining the expansion of the one-particle irreducible three-point function with that of the wavefunction renormalization, 
we obtain
\begin{eqnarray}
Q e \, V 
&=&
\left[ 1 + Z^{(1)} + Z^{(2)} \right] Q e 
+ 
\left[ 1 + Z^{(1)} \right] \Gamma^{(1)}
\notag \\
&& \phantom{space}
+ 
\Gamma^{(2)}
+ 
M^{(1)} \, 
\Gamma^{\prime (1)}
\label{eq:Gamma}
,\end{eqnarray}
which is valid up to second order in $\alpha$. 
Notice we have used the leading-order results,
$\Gamma^{(0)} = Q e$
and
$Z^{(0)} = 1$. 
Our goal is to compute 
$V$
up to next-to-leading one-loop order and leading two-loop order. 
Already from Eq.~\eqref{eq:Gamma}, 
we see that
$V^{(0)} = 1$. 
For the validity of the Ward identity, 
it thus remains to show 
$V^{(1,1)} = V^{(1,2)} = V^{(2,1)} = 0$, 
which is the desired statement of charge non-renormalization.

At the leading one-loop order, 
we have, 
on account of Eqs.~\eqref{eq:Zs} and \eqref{eq:Gamma}, 
the simple relation
\begin{equation}
Q e \, V^{(1,1)}
=
Q e \, \Sigma^{\prime (1,1)}
+ 
\Gamma^{(1,1)}
.\end{equation}
The one-loop diagram for the three-point function is shown in Fig.~\ref{f:VLONLO}. 
A straightforward computation shows 
$\Gamma^{(1,1)} = - Q e \, \Sigma^{\prime (1,1)}_\text{1-loop}$, 
where the one-loop self-energy derivative appears in Table~\ref{t:summy}. 
Consequently, 
we have 
$V^{(1,1)} =0$. 
At the next order in NRQED, 
we have similarly
\begin{equation}
Q e \, V^{(1,2)}
= 
Q e \, \Sigma^{\prime (1,2)}
+ 
\Gamma^{(1,2)}
,\end{equation}
where the next-to-leading one-loop diagrams contributing to 
$\Gamma^{(1,2)}$
are shown in Fig.~\ref{f:VLONLO}. 
It is easy to show that 
$\Gamma^{(1,2)} = - Q e \, \Sigma^{\prime (2,1)}_\text{1-loop}$, 
and one obtains the desired result
$V^{(1,2)} = 0$.

At the leading two-loop order, 
the Ward identity provides a more non-trivial check. 
On account of 
Eqs.~\eqref{eq:Zs} and \eqref{eq:Gamma},
the second-order vertex is given by
\begin{eqnarray}
Q e \, V^{(2,1)}
&=&
Q e \, Z^{(2,1)}
+ 
Z^{(1,1)} \Gamma^{(1,1)} 
\notag \\
&& \phantom{spa}
+
\Gamma^{(2,1)}
+ 
M^{(1,1)} \Gamma^{\prime (1,1)}
.\end{eqnarray}
Writing out the wavefunction renormalization factors, 
and using
$\Gamma^{(1,1)}$
determined above, 
two terms cancel and the expression simplifies to
\begin{eqnarray}
Q e \, V^{(2,1)}
&\to&
Q e \, \Sigma^{\prime (2,1)}
+ 
\Gamma^{(2,1)}
\notag \\
&& \phantom{s}
+ 
M^{(1,1)} 
\left[ Q e \, \Sigma^{\prime \prime (1,1)} + \Gamma^{\prime (1,1)} \right]
.\,\,\end{eqnarray}
Using the one-loop correction to the charge-density interaction, 
it is straightforward to show that
$\Gamma^{\prime (1,1)} = - Q e \, \Sigma^{\prime \prime (1,1)}_\text{1-loop}$, 
and thus the bracketed term vanishes.

Finally we must consider the 
$\cO(\alpha^2)$ 
diagrams contributing to the irreducible three-point function. 
There are both two-loop and one-loop diagrams, 
so we write
$\Gamma^{(2,1)} = \Gamma^{(2,1)}_\text{2-loop} + \Gamma^{(2,1)}_\text{1-loop}$,
and update the expression for the vertex to read 
\begin{eqnarray}
Q e \, V^{(2,1)}
&\to&
Q e \, \Sigma^{\prime (2,1)}_\text{2-loop}
+ 
\Gamma^{(2,1)}_\text{2-loop}
+ 
Q e \, \Sigma^{\prime (2,1)}_\text{1-loop}
+ 
\Gamma^{(2,1)}_\text{1-loop}
.\notag \\
\label{eq:almost}
\end{eqnarray}
Given the spin-independence NRQED at this order, 
the diagrams for 
$\Gamma^{(2,1)}_\text{2-loop}$
have essentially been computed above in the context of mass counterterm insertions at 
$\cO(\alpha^3)$, 
see the two-loop diagrams appearing in 
Fig.~\ref{f:Nleading3ctms}. 
We need only exert care about the sign of the result. 
The charge-density interaction appears in the Lagrangian as the term
$\cL = - Q e A_0 \psi^\dagger \psi$, 
i.e.~with the same sign as the mass counterterm
$\Delta \cL = - \delta_M \psi^\dagger \psi$. 
We conveniently define the three-point function at tree-level to be 
$\Gamma^{(0)} = Q e$, 
which represents the matrix element of 
$- \frac{\partial \cL}{\partial A_0}$,
i.e.~exactly the same sign as the contribution to the self energy from the matrix element of the mass counterterm
$- \Delta \cL$. 
It is then trivial to modify Eq.~\eqref{eq:twoloopMins} to extract the contribution to the three-point function of the charge-density interaction, 
i.e.~just replace 
$\delta_M^{(1,1)}$ with 
$Q e$. 
This produces the result
$\Gamma_\text{2-loop}^{(2,1)} = - Q e \, \Sigma_\text{2-loop}^{\prime (2,1)}$, 
and the first pair of terms appearing in Eq.~\eqref{eq:almost} exactly cancel. 
The final contribution to the 
$\cO(\alpha^2)$
charge-density interaction vertex arises from one-loop diagrams formed with an insertion of the leading-order mass counterterm, 
$\delta_M^{(1,1)}$. 
These diagrams are shown in the bottom row of 
Fig.~\ref{f:VLONLO}, 
and evaluate to 
$\Gamma_\text{1-loop}^{(2,1)} = - Q e \, \Sigma^{\prime (2,1)}_\text{1-loop}$. 
This final contribution ensures that indeed
$V^{(2,1)} = 0$.

\section{Renormalized Perturbation Theory}%
\label{s:B}%

Bare perturbation theory is conceptually more straightforward for the computation of finite-volume corrections to the electromagnetic mass. 
Renormalized perturbation theory, 
by contrast, 
makes automatic many of the cancellations encountered above. 
The leading two-loop computation is repeated here using renormalized perturbation theory, 
however, 
it exposes a subtlety in enforcing the renormalization conditions.

In terms of the renormalized NRQED matter field 
$\psi_r$, 
the Lagrangian density of renormalized perturbation theory takes the form%
\footnote{
Notice we have capitalized on the non-renormalization of electric charge to write 
$\delta_{e} = \delta_{Zr}$ 
in the renormalized Lagrangian density. 
This equality of renormalization constants was explicitly checked in 
Appendix~\ref{s:A}, 
arises as a consequence of the Ward identity, 
and does not require the imposition of a separate renormalization condition. 
}
\begin{eqnarray}
\cL_r
=
\psi_r^\dagger \, i D_0 \psi_r
+ 
\psi_r^\dagger \left( i \delta_{Zr} D_0 - \delta_{Mr} \right) \psi_r
\label{eq:Lren}
.\end{eqnarray}
Determination of the renormalization constants
($\delta_{Zr}$ and $\delta_{Mr}$) 
order-by-order in QED perturbation theory is made by enforcing renormalization conditions. 
The general behavior of the two-point correlation function has been given above in 
Eq.~\eqref{eq:2pt}. 
We adopt the following renormalization conditions on the self-energy function. 
First the mass 
$\cM_r$
is required to vanish in the infinite volume limit
\begin{eqnarray}
\lim_{ L \to \infty}
\Sigma_r(\cM_r) \equiv 0
\label{eq:masscondo}
.\end{eqnarray}
This condition mandates that the physical electromagnetic mass is already included in the large mass scale of the hadron, 
which is integrated out to give the NRQED expansion. 
The residual mass computed using NRQED will thus only be the finite-volume effect. 
The wavefunction renormalization is determined by the second condition, 
which we take to be
\begin{eqnarray}
\frac{d \Sigma_r (p_0)}{d p_0} \Big|_{p_0 = \cM_r} 
\equiv
0
.\end{eqnarray}
For great convenience, 
the latter condition is taken at finite volume.

With these two renormalization conditions, 
the pole position, 
$\cM_r = \Sigma_r (\cM_r)$,
is determined in QED perturbation theory simply as 
\begin{eqnarray}
\cM_r^{(1)} 
&=& 
\Sigma_r^{(1)}(0), 
\quad
\cM_r^{(2)} 
=
\Sigma_r^{(2)}(0), 
\quad 
\cdots
.\end{eqnarray}
To leading order, 
Eq.~\eqref{eq:Sig_1_1_p0},
we have the one-loop self-energy function
$\Sigma^{(1,1)}_\text{1-loop} (p_0)  = \Sigma_\text{1-loop}^{(1,1)} + p_0 \, \Sigma_\text{1-loop}^{\prime (1,1)} + \cdots$. 
Taking into account the counterterm Lagrangian density, 
we must enforce the condition
\begin{eqnarray}
\lim_{L \to \infty}
\Sigma_\text{1-loop}^{(1,1)} + \delta_{Mr}^{(1,1)}
=
0
,\end{eqnarray}
which yields 
$\delta_{M r}^{(1,1)} = - \frac{1}{2} (Qe)^2 \hat{\mathfrak{C}}_2$, 
as well as 
\begin{eqnarray}
\Sigma_\text{1-loop}^{\prime (1,1)} - \delta_{Zr}^{(1,1)}
=
0
,\end{eqnarray}
which produces 
$\delta_{Zr}^{(1,1)} = \frac{1}{2} (Qe)^2 \mathfrak{C}_3$. 
The leading-order finite-volume correction is determined from 
\begin{equation}
\cM^{(1,1)}_r
=
\Sigma_\text{1-loop}^{(1,1)} + \delta_{Mr}^{(1,1)}
=
\frac{1}{2} (Qe)^2 C_2
,\end{equation}
and agrees with that in Table~\ref{t:summy}.

%
%
%
\begin{figure}
\resizebox{0.8\linewidth}{!}{
\includegraphics{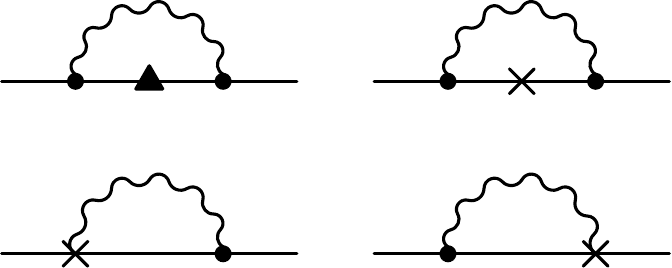}
}
\caption{
Graphical depiction of the 
$\cO(\alpha^2)$ 
one-loop diagrams required to compute the self-energy in renormalized perturbation theory. 
Diagram elements are the same as in Fig.~\ref{f:oneloop}. 
Additionally the filled triangle denotes the mass counterterm 
$\delta_{Mr}^{(1,1)}$, 
while the $\times$'s correspond to terms of the counterterm Lagrangian density proportional to 
$\delta_{Zr}^{(1,1)}$. 
}
\label{f:2ren}
\end{figure}
%
%
%

Beyond one-loop order, 
we must compute all 
$\cO(\alpha^2)$
contributions to the mass, 
and then enforce the renormalization condition in 
Eq.~\eqref{eq:masscondo}. 
Such irreducible contributions arise from 
i) two-loop diagrams formed form the leading-order vertices, 
and
ii) one-loop diagrams formed from the leading-order vertices and the insertion of one counterterm. 
Computation of the two-loop diagrams has been detailed in the main text;
and, 
from Table~\ref{t:summy},  
we have the factorized result,
$\Sigma_\text{2-loop}^{(2,1)} = - \Sigma_\text{1-loop}^{(1,1)} \Sigma_\text{1-loop}^{\prime (1,1)}$.  
The required one-loop diagrams are shown in 
Fig.~\ref{f:2ren}, 
and are straightforward to evaluate. 
The mass counterterm insertion evaluates to 
$ - \delta_{Mr}^{(1,1)} \Sigma_\text{1-loop}^{\prime(1,1)}$, 
while the 
$\delta_{Zr}^{(1,1)}$
insertion inside the loop produces the result
$- \delta_{Zr}^{(1,1)} \Sigma_\text{1-loop}^{(1,1)} $. 
The diagrams with renormalized charge vertices are trivially related to the one-loop self energy, 
namely they evaluate to
$2  \delta_{Zr}^{(1,1)} \Sigma_\text{1-loop}^{(1,1)}$.

Adding up all contributions, 
we have considerable cancellations leading to the result
\begin{eqnarray}
\Sigma^{(2,1)}_r (0)
&=&
\Sigma_\text{2-loop}^{(2,1)}
+ 
\Sigma_\text{1-loop}^{(2,1)}
+
\delta_{Mr}^{(2,1)}
\notag \\
&=&
\frac{1}{4}  (Qe)^4 \hat{\mathfrak{C}}_2 \, \mathfrak{C}_3
+
\delta_{Mr}^{(2,1)}
.\end{eqnarray}
In dimensional regularization, 
we have
$\hat{\mathfrak{C}}_2 \overset{DR}{=} 0$
and the renormalization condition in Eq.~\eqref{eq:masscondo} leads to the counterterm
$\delta_{Mr}^{(2,1)} \overset{DR}{=} 0$. 
The resulting finite-volume correction to the mass,  
$\cM^{(2,1)}_r = 0$, 
agrees with our determination using bare perturbation theory. 
In other regularization schemes,
the linearly divergent integral
$\hat{\mathfrak{C}}_2$
is non-vanishing and proportional to the ultraviolet cutoff. 
The mode sum 
$\mathfrak{C}_3$
is dimensionless, 
but depends logarithmically on the volume.  
In lattice regularization, 
for example, 
we have
$\hat{\mathfrak{C}}_2 \sim a^{-1}$, 
and
$\mathfrak{C}_3 \sim \log \frac{L}{a} + \text{finite}$, 
where 
$a$ 
is the lattice spacing and 
finite refers to terms that are finite in the infinite-volume limit.  
To enforce the renormalization condition, 
Eq.~\eqref{eq:masscondo},  
in general regularization schemes, 
we must choose
\begin{equation}
\delta_{Mr}^{(2,1)}
=
-
\frac{1}{4}  (Qe)^4 \hat{\mathfrak{C}}_2 \, \mathfrak{C}_3
\label{eq:fck}
,\end{equation} 
which also produces the result 
$\cM^{(2,1)}_r = 0$. 
The required counterterm, 
however, 
has logarithmic dependence on the volume, 
e.g.,
$\delta_{Mr}^{(2,1)} \sim a^{-1} \log \frac{L}{a}$
in lattice regularization. 
In order to have a well-defined infinite-volume limit, 
however,
such logarithmic counterterms are required. 
These are only relevant if one employs non-dimensionally regulated schemes 
using renormalized perturbation theory.

Finally, 
we can use the results of bare perturbation theory to derive the counterterms required in renormalized perturbation theory. 
Using the relation between the bare and renormalized fields,
namely
$\psi  = \sqrt{Z} \exp \big[- i \Delta M(\alpha) t \big] \, \psi_r$, 
the bare NRQED Lagrangian density,
$\cL = \psi^\dagger ( i D_0 - \delta_M ) \psi$,
becomes that of 
Eq.~\eqref{eq:Lren}, 
with the identification
\begin{equation}
\delta_{Mr} = Z \big[ \delta_M - \Delta M(\alpha) \big]
\label{eq:relation}
.\end{equation} 
Notice that 
$\Delta M(\alpha)$
is the physical electromagnetic mass, 
and 
$Z = 1  + \delta_{Zr}$
is the wavefunction renormalization factor. 
At leading order, 
Eq.~\eqref{eq:relation} leads to the renormalized mass counterterm
$\delta_{Mr}^{(1,1)} = \delta_M^{(1,1)} - \Delta M^{(1,0)} = - \frac{1}{2} (Qe)^2 \hat{\mathfrak{C}}_2$, 
which reproduces the result obtained above. 
At second order in $\alpha$, 
we have the relation
\begin{equation}
\delta_{Mr}^{(2,1)} 
= 
\delta_M^{(2,1)} - \Delta M^{(2,0)} 
+
\delta_{Zr}^{(1,1)} 
\left( \delta_M^{(1,1)}  - \Delta M^{(1,0)} \right)
,\end{equation}
which evaluates exactly to that given in  
Eq.~\eqref{eq:fck}.

\bibliography{bibly}

\end{document}